\documentclass[preprint,one column]{revtex4}
\usepackage{epsfig} 
\usepackage{amsmath,amssymb}
\usepackage{graphicx}

\newcommand{\Omk}{\Omega_{k0}}
\newcommand{\Om}{\Omega_{m0}}
\newcommand{\vecal}{\vec{\alpha}}

\newcommand{\beq}{\begin{equation}}
\newcommand{\eeq}{\end{equation}}

\newcommand{\wt}{w_{tot}}

\newcommand{\rt}{\rho_{tot}}

\begin{document}

\title{Do Type-Ia Supernovae Constrain the Total Equation of State?}
\author{William Komp}
\email{w0komp01@louisville.edu}
\affiliation{Department of Physics, University of Louisville, 102 Natural
  Sciences, Louisville, KY. 40292 USA}

\begin{abstract}
In this paper, we consider a couple of alternative dark energy models using
the total equation of state of the cosmological fluid, $\wt$.  These models
are fit to the recent type-Ia supernovae data and are compared to previously 
considered models.  The first model is based on the hyperbolic tangent and
provides a good estimate of the rate of the transition to dark energy
domination.  The second model is a cubic spline model.    This model
demonstrates and quantifies the non-monotonicity in the total equation of
state coming from the supernovae observations.  
At present, the supernovae observations indicate significance to
non-monotonically decreasing dark energy.  We derive constraints on the
spline paramters and compare and constrast the results to the Cosmological
Constant dark energy model. Both the hyperbolic and splines models indicate
that a precise physical notion of dark enegy is a potentially  ever more
mysterious quantity?  
\end{abstract}

\maketitle 

\section{Introduction} \label{sec:intro}

Observations of distant type-Ia supernovae have shed great light on
the evolution of universe at
late times\cite{P98,k03,Riess98,Riess01,Riess04,R06}. 
These observations indicate that the universe is presently undergoing
an accelerated expansion that started about 7 billion years ago.  The
source of this acceleration has been labeled dark energy and 
the corresponding energy density 
is peculiar in that it appears to require a negative equation of
state\cite{Staro03}.  Cosmologists were quick to pursue models which
reproduce the 
observed effects of the dark energy.  All of these models have
negative equations of state coming from positive dark energy density and 
negative pressure.  

The first of these models has its historical origins with Einstein.  It is the
Cosmological Constant Cold Dark Matter (LCDM) cosmological model.  This model
adds an additional constant $\Lambda$ to the Einstein Equations.
The resulting model of dark energy is
typically  characterized by a constant equation of state
$w_{\Lambda}\equiv P_{\Lambda}/\rho_{\Lambda}=-1$.  Initially, the
LCDM model fit observational data quite well\cite{Riess98,P98}.
More recent data suggests that while this model is not
ruled out, it is statistically not as favored as models with smaller
constant equations of state\cite{Riess04,Caldwell02}.  
Still, others propose to introduce
phenomenological models of the dark energy equation of state (and
consequently the energy density) based on kinematics\cite{Gong04} or
linear Taylor's Series expansions of the dark energy equation of state
($w_V$) \cite{Turner02,Linder03,Riess04}.  
The latter model was shown to be
divergent at early times and has since fallen out of favor.  However,
this inspired others to introduce alternative evolving dark energy
equations of state which removed the early time exponentially divergent 
behavior\cite{Linder05,padman05}.  The 
first example of these models was originally proposed by
\cite{Chevalier00} as an alternative to constant equation of state
models and has an evolving equation of state.   
In \cite{padman03}, they use this model to analyze different
properties of the supernovae data and implications for the dark
energy parameters.  For dark energy, they find that it is difficult to
tightly constrain several dark energy parameters using the recent
supernovae data.  

In \cite{Staro03}, they propose several ansatz models for the hubble
parameter which are polynomials of the cosmological scale factor.  
In these models, the dark energy is characterized by an equation of
state that metamorphosizes from $w_V=0$ at a redshift $z>1$ to $w_V
\approx -1$ at $z=0$.
From this analysis, they conclude that the supernovae data favors
negative evolving equations of state for dark energy.  
Similar ansatz models for dark energy were proposed in \cite{Odintsov}
and are consistent with the results from \cite{Staro03}.  
Still others have proposed modeling dark energy equation of state by
using cubic splines\cite{Wang04}.  
Here, they constrain dark energy in a
series models and find that they are consistent with negative equations 
of state using supernovae, gravitational lensing and large scale
structure.  The results of their analysis put tight constraints on
dark energy equations of state.  

A very curious set of models are the Sudden Gravitational Transition
models, see \cite{Caldwell05} and references therein.  These
models suppose that dark energy is the result of a late-time phase
transition in the universe.  The source of the dark energy is the 
result of a resonant effect of 
a free, ultra-low mass, quantized scalar field coming from Quantum
Field Theory in Curved Spacetime.  With a suitably
chosen order parameter $\eta$, this resonance causes
$\eta \rightarrow \chi^2 m^4={\rm constant}$ where
$\eta$ takes the form $R^2$, $R_{\mu \nu} R^{\mu
  \nu}$ and $R_{\mu \tau \nu \rho} R^{\mu \tau \nu \rho}$.  
The model for which $\eta=R^2$ is called the Vacuum Cold Dark Matter (VCDM)
cosmological model\cite{Parker99a,Parker99b,Parker00,Parker01,Parker04}.
In all of the Sudden Transition models, the effects of the scalar field are
negligible until about half the age of the present universe, denoted in terms
of redshift as $z_j$ and is labeled the redshift of transition.  
Here $\chi^2 m^4$ is the one free parameter of these
models with $\chi$ corresponding to a dimensionless parameter fixed by the
theory and $m$ is the mass of the scalar field.  The constant $\eta$
arising from this effect reacts back on the universe which 
results in an accelerated expansion through a sort of gravitational
Lenz's law.  It has been shown that these types of models are tightly constrained
by the recent cosmological experimental
data, see \cite{Caldwell05,komp04,ckpv2005,pkv2003} for details.  

The effects of these models are different than the previous ones in
that they change the nature of gravity itself.  Dark energy is not an
unknown extra quantity that appears in the total energy density 
$\rho_{tot}$ but is a manifestation of the changes in gravity induced
by the effects of the field.  However, these changes lead to
comparable results to the previous dark energy
models\cite{ckpv2005,Caldwell05,komp04}.   

A similar model to the Sudden Gravitational Transition models
is the Super-Acceleration model\cite{onemli02,onemli04}.  
This model is based on Quantum Field Theory
in Curved Spacetime as well.  A fully renormalized energy density and
pressure are computed for a scalar field with a quartic
self-interaction in a locally De Sitter spacetime.  The resulting
cosmology leads to a dark energy term in the Einstein equations which
has a dark energy equation of state which relaxes gradually from 0 to
-1.    

What we propose here is that we model directly the Hubble parameter's
evolution through the total equation of state $w_{tot}$.  This is
similar to the method from  \cite{Staro03} mentioned above.  This
approach has a 
novelty that it does not have the present day matter density $\Om$ 
(or alternatively the present day dark energy density $\Omega_{X0}$)
as a free parameter and models directly the kinematics contained in
the supernovae data, see \cite{Staro03,padman03} for nice discussions.  
We will take two approaches to the modeling.  The first approach is based on a 
phenomenological model that was inspired by the hyperbolic tangent, hence
forth labeled the hyperbolic model.
This model has two parameters which determine the time of transition
($\alpha$) and the strength of transition ($\beta$).  
This 2 parameter model is fit to supernovae data using a finite linear
grid and marginal parameter estimates are extracted using the
procedure discussed in \cite{komp04}.  
When suitably tuned by this fit, these parameters cause
$w_{tot}$ to undergo a transition that
will cause it to deviate from 0 at a redshift $z \approx 1$.  
This transition is a
monotonically decreasing function of time bounded below by -1 or conversely a 
monotonically increasing function of redshift bounded above by 0.  

For simplicity, we have set the
amplitude of the transition to unity.  This sets the cosmology to asymptote to
de Sitter spacetime, i.e. 
$w_{tot}$ asymptotes to $-1$.  This is motivated from the results
presented in \cite{Wang04}, where they show that the
supernovae data does not constrain the dark energy density for 
redshifts $z<0$.  If some data in the future becomes available to
constrain the dark energy at these redshifts, then it is
possible to adjust the amplitude and it would become an additional
parameter.  We will show that the transition in this model is very
reminiscent of the models of dark energy discussed above and in
particular behaves most like the Sudden Gravitational Models and the
dark fluid model\cite{Arbey05}.  

The second approach to modeling $w_{tot}$ that we will take is a
phenomenological cubic spline analysis.  Our interests are to  constrain cubic 
spline models of $\wt$ to the ranges of redshift associated with dark energy  
domination between redshifts ($0<z<1$).  We choose two different
uniform densities for the splines, consisting of 3 and 6 points.  
The spline points correspond to 
the model parameters and will be denoted
as $a_i$ for $i=1,2,3$ for the 3-point model or $i=1,...,6$ for 
the 6-point model. 


We choose to vary the spline points over a finite uniform linear grid.  
Performing the fit to the supernovae data at each point.  
From this analysis, we conclude that the supernovae data favors not just
monotonically decreasing $\wt$ with one extremal point, 
but allows (and marginally favors for some data sets) transitions which 
have several extremal points.  However, implications for
monotonically decreasing $\wt$ shows that the results of the various
proposed models above are within the constraints allowed by the
spline analysis.  

Both of the cosmologies proposed in this paper have an 
earlier matter dominated epoch followed by a later dark energy
dominated epoch (i.e. $\wt=0$).  In the earlier matter dominated epoch, the
energy density which would characterize this later epoch appears 
as (presumably) cold (presumably) dark matter like some of the models
considered in \cite{Staro03}.  Thus, these models have some subset
of dark matter which undergoes a transition to dark energy.  This
transition is similar to the dark fluid model presented in
\cite{Arbey05}.  There, he supposes that the riddle of dark energy and
dark matter are one in the same.  Dark Matter begins to undergo some
transition, like a decay, into dark energy at $z \approx 1$.  

In section \ref{sec:phenmodel}, we will propose
the two models of $\wt$ and fit them to the recent supernovae data.  One is
a toy model based on the hyperbolic tangent and the other is a cubic spline model.  
In section \ref{sec:disc}, we will discuss comparison of the models considered
in the previous section to the LCDM model.  Also, comparison of the models to
the recent WMAP data through the shift parameter will be discussed.  


In section \ref{sec:con}, we
will summarize our results and make concluding remarks about these two models
and compare them to the LCDM model.  
It should be noted that implicit in this discussion is the 
Friedmann-Robertson-Walker (FRW) spacetime invariant line element
given by 
\beq \label{eq:frwline}
ds^2=-dt^2+a^2(t) \left( \frac{dr^2}{1-kr^2}+r^2 (d\theta^2+\sin^2(\theta)
  d\phi^2 )\right)\;, 
\eeq
where $k$ specifies the curvature of the spatial hypersurfaces in spacetime.
In this paper, we will assume spatially flat hypersurfaces, i.e. $k=0$.  Also,
we assume that the energy-momentum-stress tensor is given by a perfect fluid with
total energy density ($\rho_{tot}$) and pressure ($P_{tot}$).

\section{Modeling Dark Energy} \label{sec:phenmodel}
 
Typically, cosmologists interpret the observed late-time acceleration as arising
out of some non-standard cosmological energy density.  In order for this energy
density to achieve the observed acceleration, it needs to exert a significant negative
pressure at late times in the universe's evolution.  This means that
the ratio of pressure to energy density 
(presumed to be non-negative) is negative, i.e. the dark energy equation of
state $w_V(z)=P_V/\rho_V<0$.  All reasonable fitting dark energy models
are in agreement on this point.  For the LCDM model, $w_V(z)$ is equal to
$-1$ for all redshifts.  For the VCDM model and the other sudden gravitational
transition model
\cite{pkv2003,komp04,Caldwell05}, $w_V(z)$ 
is shown to be less than $-1$ for late-times after the transition to
dark energy domination.  Similarly in \cite{Linder05,Wang04,padman05}, 
they consider a
significant number of dark energy models and find $w_V(0)<0$ is favored.  

With this in mind, we propose to model dark energy with no
explicit dependence on $\Om$ by using the total
equation of state of the universe $w_{tot}$ given by

\beq \label{eq:eq1}
w_{tot}=\frac{P_{tot}(z)} {\rho_{tot}(z)}\;,
\eeq  

where $P_{tot}(z)$ and $\rho_{tot}(z)$ are the total pressure and energy
density of the universe respectively.  The benefit of this approach of 
modeling the effects of dark energy is that at late times any
parameters introduced into \eqref{eq:eq1} which cause $\wt$ to become 
negative correspond to dark energy parameters.  
Previously, one had the parameter $\Om$
with which to contend and when combined with other data would put
constraints on fitting to the supernovae observations. 
We will constrain the empirically permissible deviations from a Standard Cold
Dark Matter model.
It should be noted that one could assume this approach and arrive at
all of the expressions for $w_V(z)$ arising out of the various dark
energy models.  So, this approach changes nothing for the previous
models but allows for a new angle of examination of the supernovae
observations which can be compared and contrasted to models that have 
already been examined.  

Now, assuming the invariant line element in \eqref{eq:frwline},
then the time-time Einstein Equation is given by
\beq \label{eq:phen00ee}
\frac{H^2(z)}{H^2_0}=\frac{8 \pi G}{3 H_0^2} \rho_{tot,0}
\frac{\rho_{tot}(z)}{\rho_{tot,0}}\;,
\eeq
where $H^2_0$ is the present day hubble constant and $\rho_{tot,0}$ is
the present day value of the total energy density.  We are
assuming a spatially flat universe.  Thus,the total energy density $\Omega_{0}
\equiv (8 \pi G)/(3 H_0^2) \rho_{tot,0}=1$.  Substituting this result into
\eqref{eq:phen00ee} gives  

\beq \label{eq:phen00eea}
\frac{H^2(z)}{H^2_0}=\frac{\rho_{tot}(z)}{\rho_{tot,0}}\;.
\eeq

Assuming $T^{\mu \nu}$ is a perfect fluid, $\nabla_{\mu}T^{\mu \nu}=0$ implies
that $\frac{\rho_{tot}(z)}{\rho_{tot,0}}$ satisfies the following expression 

\beq \label{eq:rhotot}
\frac{\rho_{tot}(z)}{\rho_{tot,0}}= 
{\rm Exp} \left[ 3 \int_0^z \frac{dz'}{1+z'}
 (1+w_{tot}(z')) \right] \;,
\eeq
where we have explicitly written the redshift dependence of $\wt$.  

From this equation, we can see that modeling $\wt$ is directly
modeling the hubble parameter, \eqref{eq:phen00eea}, which is
the method used in \cite{Staro03}.  So in this paper, we are
presenting alternative models of the hubble parameter.  

\subsection{The Hyperbolic Model} \label{sec:phenmodela}
Now, lets suppose a particular phenomenological model of $w_{tot}$
which will characterize the transition to dark energy domination.  This model
is inspired by the hyperbolic tangent function and will be labeled the
hyperbolic model.  We will assume a prior matter dominated epoch for $z
\approx 1$ and above, i.e. $w_{tot}=0$.  As discussed above in order to get
a dark energy model with late-time negative pressure, $w_{tot}$ must
become negative as $z$ goes to 0 (present day).  Let $w_{tot}$ be defined by   
\beq \label{eq:phenmodel}
w_{tot}=-\frac{{\rm Exp}[- \beta (z-\alpha)]} {{\rm Exp}[- \beta
    (z-\alpha)]+{\rm Exp}[+ \beta (z-\alpha)]}\;,
\eeq
where $\beta$ and $\alpha$ are labeled as transition parameters.
$\beta$ controls the strength of the transition to dark energy
dominated epoch and $\alpha$ controls the time of transition.  

Substituting \eqref{eq:phenmodel} into \eqref{eq:rhotot}, we get  
\begin{widetext} 
\beq \label{eq:phenomratrho}
\frac{\rho_{tot}(z)}{\rho_{tot,0}}={\rm Exp}\left[3 \int_{0}^{z}
  \frac{dz'}{1+z'} \left( 1+\frac{{\rm Exp}[- \beta (z'-\alpha)]}
       {{\rm Exp}[- \beta
    (z'-\alpha)]+{\rm Exp}[+ \beta (z'-\alpha)]} \right) \right]\;.
\eeq 
\end{widetext}
Substituting this expression into the time-time Einstein equation
\eqref{eq:phen00eea} gives
the following expression for the hubble parameter for this
phenomenological model 
\begin{widetext}
\beq \label{eq:phenhubble}
\frac{H^2(z)}{H^2_0}={\rm Exp} \left[ 3 \int_{0}^{z}
  \frac{dz'}{1+z'} \left( 1+\frac{{\rm Exp}[- \beta (z'-\alpha)]} {{\rm Exp}[- \beta
    (z'-\alpha)]+{\rm Exp}[+ \beta (z'-\alpha)]} \right) \right] \;.
\eeq
\end{widetext}
Now, we fit to the experimental data from R04 and SNLS through the luminosity
distance following the $\chi^2$ procedure given in \cite{ckpv2005,komp04} using
\eqref{eq:phenhubble} to give the hubble parameter as a
function of redshift.  

\begin{figure}
$
\begin{array}{ccc} 
\includegraphics[width=2.0in]{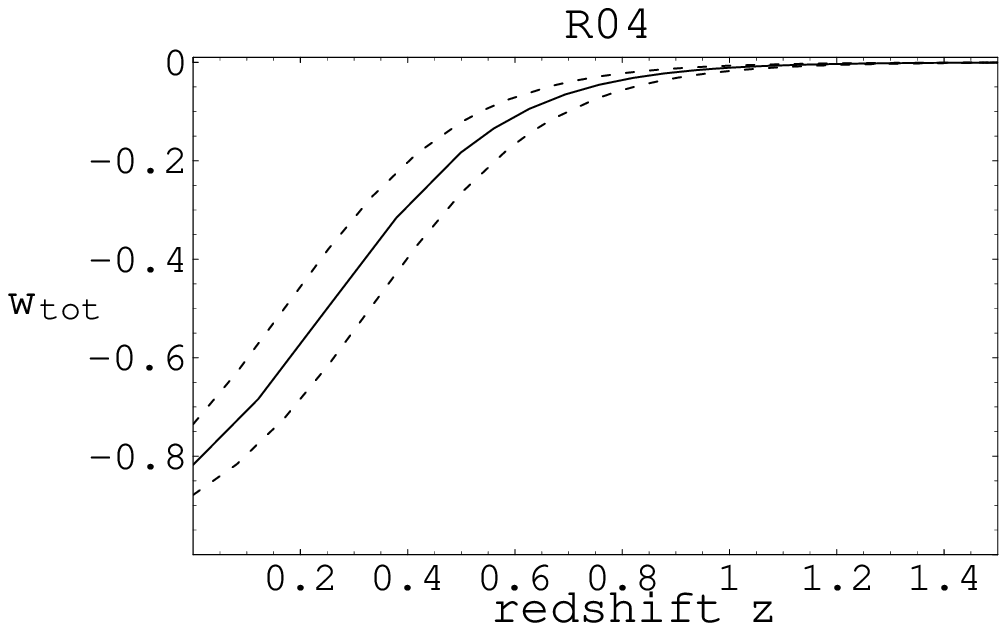} & 
\includegraphics[width=2.0in]{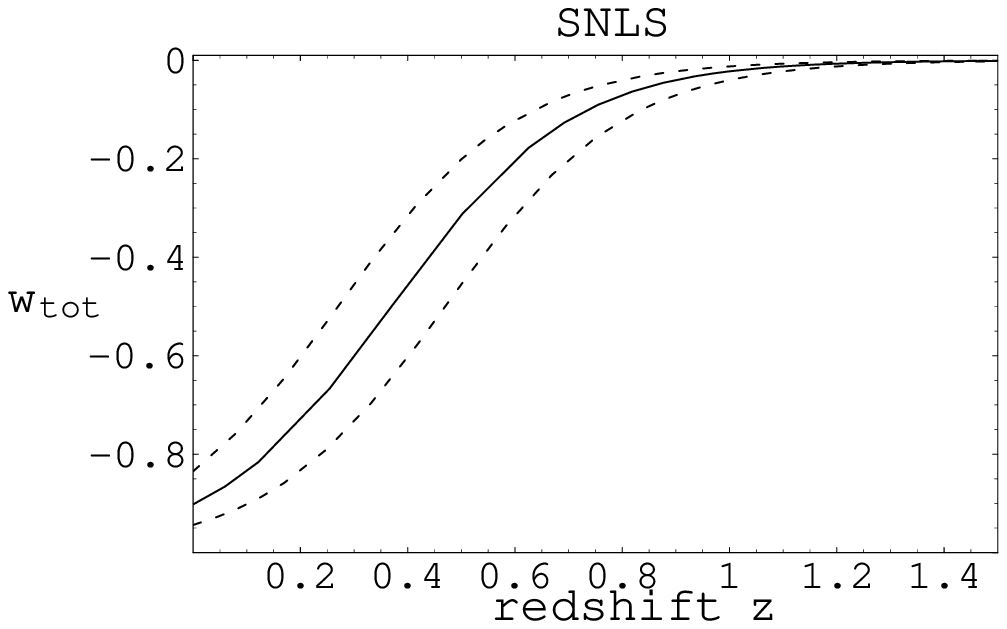} &
\includegraphics[width=2.0in]{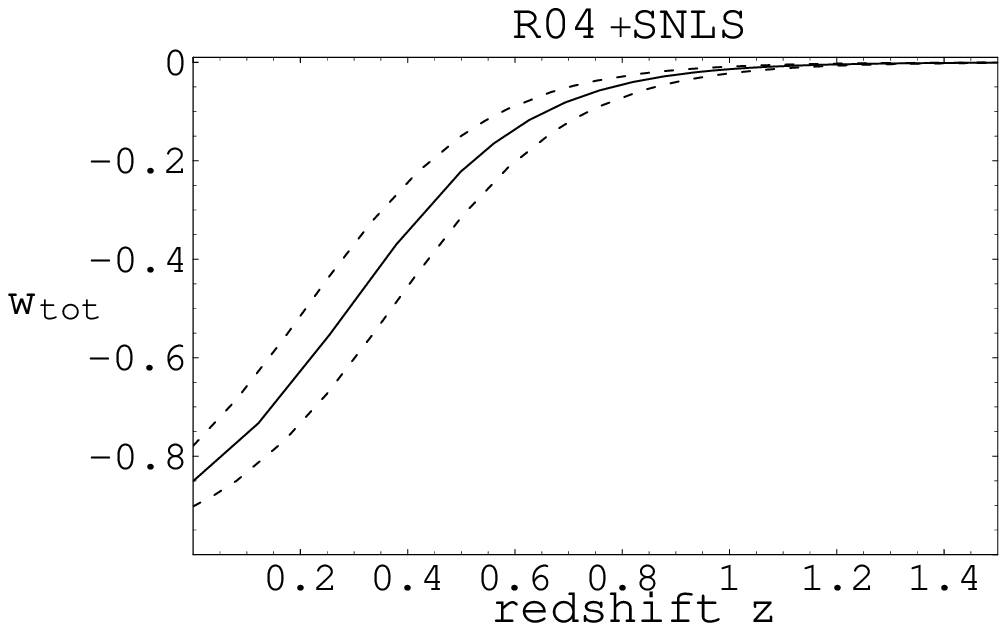} \\
\includegraphics[width=2.0in]{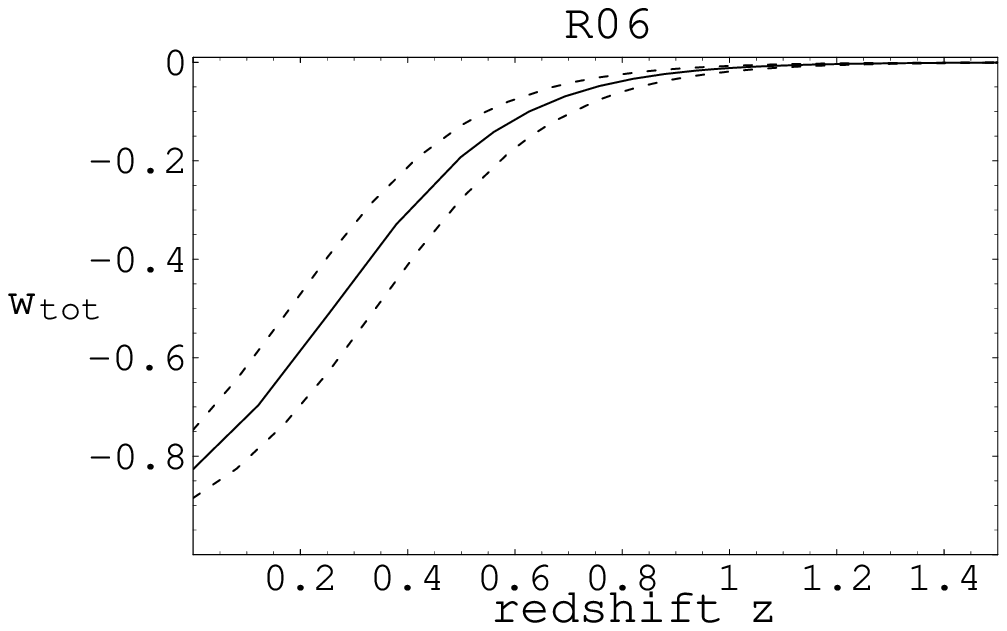} & 
\includegraphics[width=2.0in]{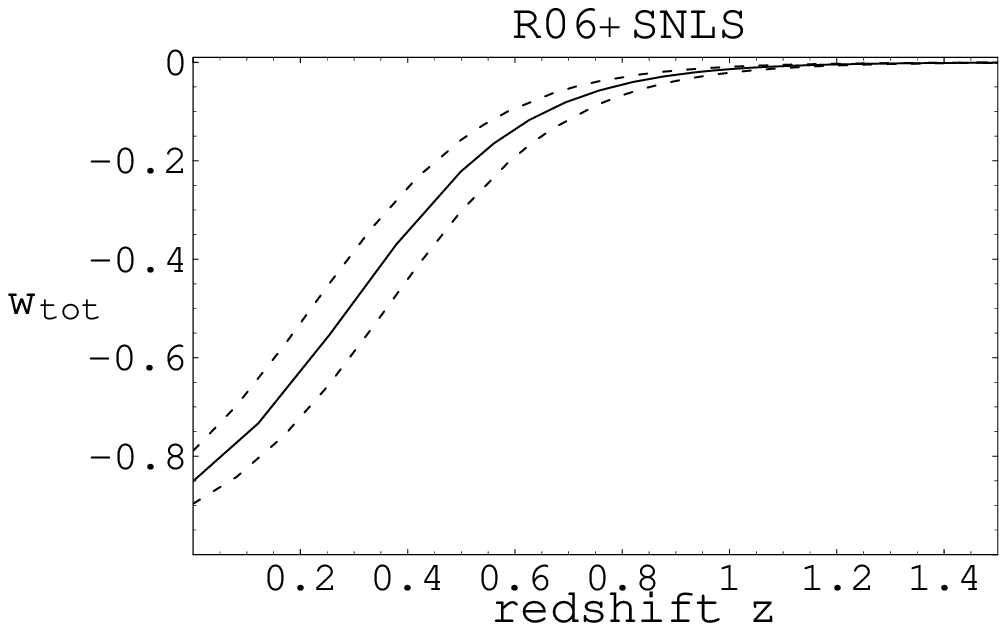} & 
\end{array}$
\caption{Plots of the $w_{tot}$ and 2$\sigma$ confidence region 
  for the hyperbolic dark energy model given by \eqref{eq:phenmodel} for the
  supernovae data sets R04, R06, SNLS, R04+SNLS and R06+SNLS data sets.  The
  solid curves in each graph correspond to the marginal estimate for $\alpha$
  for each data set presented in table \ref{tab:hypresults}.  
  The dashed lines correspond to the uncertainty in $\alpha$. For these
  graphs, we set $\beta=3$ which is near the center of the ranges given in the
  table.  From the graph, this model starts out asymptotically in a matter
  dominated stage,$\wt=0$.  As $z \rightarrow 1$, the dark energy becomes
  significant driving $\wt$ negative.  This model by construction asymptotes
  to a LCDM model as $z \rightarrow -1$.
\label{fig:phenwtot}}
\end{figure}

Performing this fit, we get the results presented in table
\ref{tab:hypresults}.  The first column corresponds to the data set(s) used in
the fitting.  The
second column corresponds to the local minimum in $\chi^2$.  The last two
columns correspond to the marginal parameter estimates and uncertainties of
$\alpha$ and $\beta$.  The resulting fits for each data set are marginally better
than for the LCDM model.  Also, the transition and evolution of the dark
energy appears to favor a greater negative pressure consistent than that which
the LCDM model favors.  This is shown for each data set in figure
\ref{fig:phenwtot}. 

From this figure, the transition to dark energy domination ($z_j$) occurs at $z
\approx 1$ and is smooth like the LCDM.    
From the fit to the supernovae data, this redshift
is smaller than the corresponding redshift of transition for the LCDM
model. However, the hyperbolic dark energy increases 
very rapidly and quickly dominates the energy density in the Einstein
Equation \eqref{eq:phen00eea}.  
So, while the hyperbolic dark energy model near $z_j$ is smooth like
the LCDM, the transition of the dark energy is more akin to the Sudden
Gravitational Transition\cite{Caldwell05} and Dark Fluid Model\cite{Arbey05}.    

\begin{table}
\caption{\label{tab:hypresults} A table of the marginal parameter estimates of
the hyperbolic model derived from the R04, R06 and SNLS data at $2\sigma$.}
\begin{ruledtabular}
\begin{tabular}{ccccc}
Data Set      & $\chi^2_{{\rm min}}$ & $\alpha$ & $\beta$  \\
\hline
R04   &  175 & $0.25 \pm 0.08$ &  $7.0^{+8.0}_{-7.0}$  \\
SNLS  & 62 & $0.37 \pm 0.10$ &  $2.1^{+2.8}_{-2.0}$  \\
R04+SNLS  & 238 & $0.29 \pm 0.08$ & $3.4 \pm 3.2$  \\
R06 & 201 & $0.26 \pm 0.08$ & $6.3 \pm 5.0$ \\
R06+SNLS & 264 & $0.29 \pm 0.07$ & $3.2 \pm 2.8$ 
\end{tabular}
\end{ruledtabular}
\end{table}

From table \ref{tab:hypresults}, the $\alpha$ parameter is constrained and 
$\beta$ appears weakly constrained.  However, the trend from this table is
that the most recent supernovae observations are beginning to provide
additional information which results in tighter constraints on cosmological
parameters.  Thus, the supernovae data only constrains
one of our transition parameters.  In figure \ref{fig:phenwtot}, we see a plot
of the 2$\sigma$ confidence region for $\wt$ coming from the marginal
estimates and uncertainties.  This is in agreement with
\cite{Linder05,Wang04,Gong04,Caldwell05,komp04}, 
where they show from the analysis of many models that only one
parameter of dark 
energy is constrained by the data.  This also agrees with the results obtained
below in section \ref{sec:disc}.  

In \cite{Chevalier00,Linder05}, their analysis assumed 
particular forms of the dark energy density ($\rho_V(z)/ \rho_{V}(0)$)
with the constrained parameters roughly corresponding to the dark
energy equation of state $w_V(z=0)$ and $w'_V(z=0)$.  One of the simplest
model is of the form
\beq \label{eq:linder20}
w_V(z)=w_0+w_1 \frac{z}{1+z}\;,
\eeq
where $w_0$ and $w_1$ are both parameters of the
model\cite{Chevalier00}.  This model has been labeled 
as model 2.0 in \cite{Linder05}.  We will assume this label here.  

In \cite{Wang04}, they used spline approximations of the dark energy
equation of state.  We will consider a similar model below.  In
\cite{Gong04}, he assumed a kinematic expansion 
of the cosmological scale factor up to the snap term ($\propto
\ddddot{a(t)}$).  The Sudden Gravitational
Transition models from \cite{Caldwell05} assumed a spatially flat
universe where the one free model parameter 
corresponds to $\Om$.  This parameter was found to be tightly
constrained.  However, if one introduces curvature, $\Om$ is found to
have a high degree of covariance with $\Omk$ which is not tightly
constrained\cite{komp04}.  This result is similar to that shown in
figure 4 of \cite{padman03}.  

\begin{figure}
\includegraphics[width=3in]{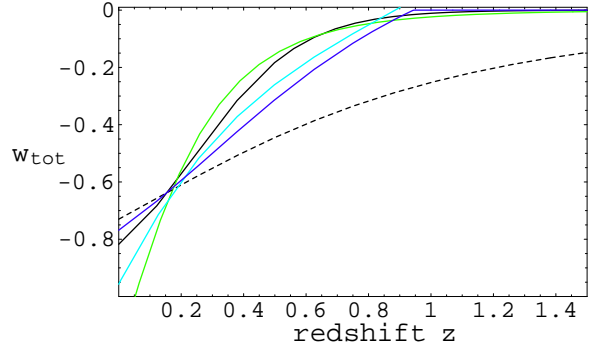}
\caption{(color online) A comparison plot of $w_{tot}$ for various dark energy
  models.  The light blue curve corresponds to the linear Taylor's
  Series model from R04, with parameters $\Om=0.27$, $w_0=-1.3$ and
  $w'_0=1.5$. 
  The green curve corresponds to the Linder 2.0 model (defined by
  \eqref{eq:linder20}) with parameters
  $w_0=-2.25$ and $w_1=0.3$.  
  The purple curve corresponds to the N=4 VCDM model with
  $\Om=0.45$.  The dashed curve corresponds to the LCDM model with
  $\Om=0.27$.  The solid black curve is the phenomenological model
  presented in this paper with $\alpha=0.25$ and $\beta=3$.  
\label{fig:compplot}}
\end{figure}

A comparison plot of several of these models and the hyperbolic 
one is given in figure \ref{fig:compplot}.  
From this figure, we see that all of the models appear to converge in
their prediction of $w_{tot}$ at $z \approx 0.19$.  Thus at this
redshift, the supernovae
predicts that $w_{tot} \approx -0.6$.  Notice that the
Linder 2.0 model and the Linear Taylor's Series model have
divergences.  The former diverging in the distant future of $z=-1$ and
the latter diverging at $z=-1$ and $z \rightarrow \infty$.  The
hyperbolic model proposed here has no such divergences and most
closely resembles the VCDM model (\cite{Parker04} and references
therein) and the dark fluid model of \cite{Arbey05}.  

\begin{figure}
\includegraphics[width=3in]{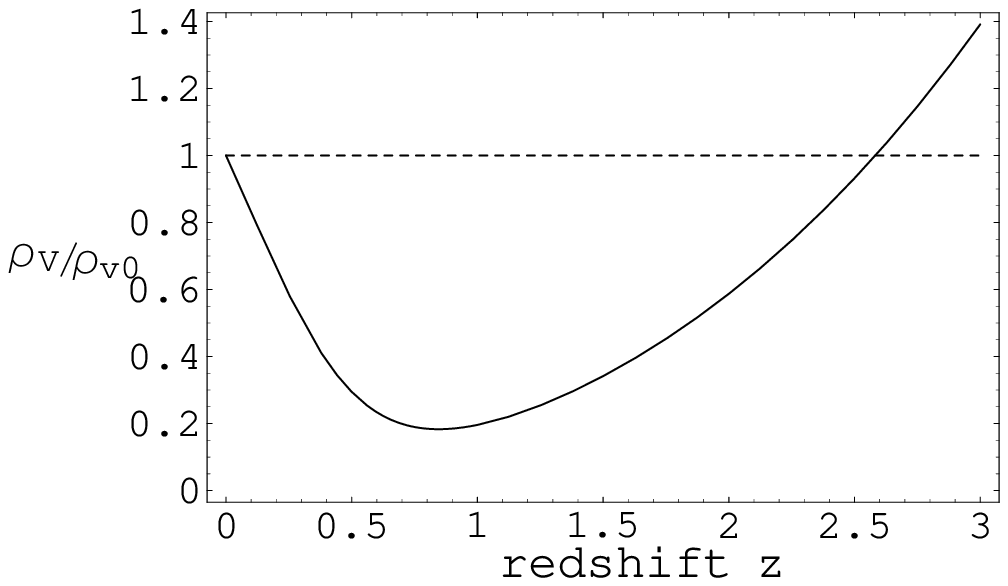}
\includegraphics[width=3in]{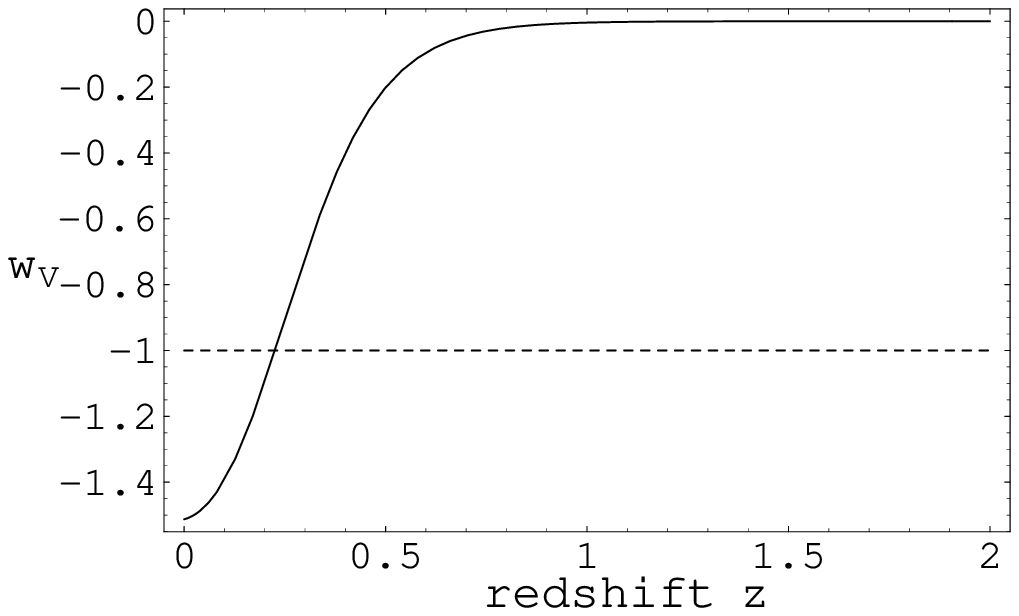}
\caption{A comparison plot of the predicted ratio dark energy density to its present
  day value versus redshift (upper) and the dark energy equation of
  state (lower) for the phenomenological model with $\alpha=0.29$ and
  $\beta=3.0$ assuming $\Om=0.45$ (solid curves) and the LCDM model (dashed
  curves).  
  At increasing redshift this model indicates that the energy density of 
  the universe is behaving identical to that of non-relativistic matter.  
  Approaching present day, this decreasing energy density reaches a minimum 
  a begins to increase again.  This is reminiscent to the effects which occur 
  in the Sudden Gravitational Transition models\cite{Caldwell05},
  the dark fluid model proposed in
  \cite{Arbey05} and the ansatz models from \cite{Staro03}. Clearly, the data
  allows for physical models which have significantly different dynamics.
\label{fig:rhovplot}}
\end{figure}

From the early redshift behavior, it would appear that this model
suffers from a similar divergence problem as the linear Taylor's
Series model 
used in \cite{Riess04,komp04}, but this is not the case.  Upon inspection of
\eqref{eq:rhotot}, one sees that $w_{tot}$ vanishes for $z
\approx 1$ and greater.  This means that during the preceding matter
dominated stage, this form of dark energy behaves just like
pressureless non-relativistic matter, i.e. grows as $(1+z)^3$ and has
an equation of state $w_V=0$.  Thus, this dark energy model is some
form of the dark fluid from \cite{Arbey05} and the ansatz models from
\cite{Staro03}.  
For $z>1$, $\Omega_{V}(z) (\equiv 8 \pi G \rho_V(z)/(3 H_0^2))$ is given
by 
\beq
\Omega_V(z) \approx \left( \Omega_{Vj}-\Om \right) (1+z)^3\;,
\eeq
where $\Omega_{Vj}$ is some constant determined by $w_{tot}$ at the
redshift of transition ($z_j$) which is a function of $\beta$ and
$\alpha$.  For typical values of these two parameters coming from
fitting to the supernovae data (below), $\Omega_{Vj}\approx 1.5 \Om$.  
Thus at early time, this form of dark energy would only make the
universe appear that it has more cold matter than it really does.  This is
the essence of the dark fluid model where dark energy arises out of
dark matter around the transition at $z=z_j$.
We can compare the hyperbolic model to the other dark energy models by
introducing the assumption that $\rho_{tot}=\rho_{m}+\rho_V$ where $\rho_m$
and $\rho_V$ are the energy densities of matter and dark energy
respectively. 
Upon inspection of \eqref{eq:phenmodel}
and \eqref{eq:rhotot}, $\rho_V$ for the hyperbolic model is present at all 
stages of the universe's evolution, which is a trait that it has in common
with the LCDM model.  
To make a quantitative accessment, it is necessary to assume a value for the
ratio of the matter energy density to critical density at present day, $\Om
\equiv 8 \pi G \rho_{m0} / (3 H_0^2)$.  Since the hyperbolic model most
closely resembles the VCDM model, we will assume that $\Om=0.45$ which is the
marginal estimate from that model.   
As shown in figure \ref{fig:rhovplot}, the dark energy density associated
with the hyperbolic model decreases as z approaches 1 from above behaving as
matter, $\rho_V \propto (1+z)^3$.  At
$z \approx 1$, the dark energy evolution begins to deviate from $(1+z)^3$ and
begins to grow in significance.  This is where the negative pressure from the
dark energy begins to grow resulting in the observed acceleration.  
This is indicated by the increase at low
redshift in the figure.
The behavior of this equation of state is reminiscent of the Sudden
Gravitational 
model with a non-zero cosmological constant\cite{Caldwell05} and the
dark fluid model \cite{Arbey05}.    

In this section, we have shown that the hyperbolic model fits to
the supernovae data quite well as compared to the other cosmological models.
In the next section, we generalize the approach of using $w_{tot}$ by
using spline approximations, similar to those used in \cite{Wang04}.
We have fixed by construction that this model asymptote to a LCDM
model as $z$ goes to $-1$.  However, one could introduce a non-unit
amplitude in the numerator of \eqref{eq:phenmodel} and have
it asymptote to any value.  However, introducing this parameter would
increase this model to a 3 parameter model.  Given that the supernovae
data constrains only one of the two parameters, introducing a third
would most likely not the statistical significance of the model to the
supernovae data.

\subsection{Spline Approximation}
In this section, we will use cubic splines to approximate $w_{tot}$.  
This is similar to the analysis performed in \cite{Wang04}.
There they modeled the dark energy equation of state $w_V(z)$ 
using cubic splines.  The analysis in this paper will 
assume a prior matter dominated stage ending at $z \approx 1$.  Thus, we will
assume that the spline points of $w_{tot}$ for redshifts $z>1$ vanish.  
The model parameters of these spline models are the spline points that specify
the spline functions for $z \le 1$.  
Recall from the hyperbolic and other dark energy models considered in section
\ref{sec:phenmodela}, we found that the dark energy effects become significant
near the redshift of transition, $z_j$.  We want to use the splines to try and
constrain what types of transition and evolution of the dark energy from
$z_j$($\approx 1$) to present day.  Thus, the redshifts of the parameters of
the spline model will be contained in the redshift interval [0,1].  In this
interval, we will analyze uniform grids of spline points.  Consider two
densities of spline points, 3-point and 6-point, which will be denoted as
$a_i$ for i=1-3 or 1-6.  Our main interest here is constraining $w_{tot}$
using the supernovae data, inferring properties of $w_{tot}$ and comparing the
results to other models. We will show that the computed confidence regions of
the splines models  will have significant intersection with most of those from
the dark energy models discussed in section \ref{sec:phenmodela}.  

Assuming that $\wt$ is given by a cubic spline function, we can fit the resulting
cosmological models using the same $\chi^2$ procedure that was used for the
hyperbolic model.   
Fitting to the data, we find the minimum $\chi^2$ and marginal estimates for
the 3 point spline is present in table \ref{tab:3ptmarg} and for the 6 point spline
is present in table \ref{tab:6ptmarg}.  
\begin{table}
\caption{\label{tab:3ptmarg} Marginal Parameter estimates at 2$\sigma$ and the
  minimum $\chi^2$ for the 3 point spline model with parameters $a_1$, $a_2$
  and $a_3$.}
\begin{ruledtabular}
\begin{tabular}{ccccc}
Data Set      & $\chi^2_{{\rm min}}$ & $a_1$ & $a_2$ & $a_3$ \\
\hline
R04   &  174 & $-0.5^{+0.5}_{-0.6}$ &  $-0.4 \pm 0.2$ & $-1.3 \pm 0.5$ \\
SNLS  & 62 & $-0.3^{+0.3}_{-0.4}$ &  $-0.6 \pm 0.1$ & $-0.7 \pm 0.4$ \\
R04+SNLS  & 238 & $-0.4 \pm 0.2$ & $-0.6 \pm 0.1$ & $-0.8 \pm 0.3$ \\
R06   & 200  & $-0.5 \pm 0.2$ & $-0.4 \pm 0.2$ & $-1.3 \pm 0.5$ \\
R06+SNLS & 262 & $-0.4 \pm 0.2$  & $-0.6 \pm 0.1$  & $-0.7 \pm 0.4$ \\
\end{tabular}
\end{ruledtabular}
\end{table}
\begin{table}
\caption{\label{tab:6ptmarg} Marginal Parameter estimates at 2$\sigma$ and the
  minimum $\chi^2$ for the 6 point spline model with parameters $a_1$, $a_2$, $a_3$,
  $a_4$, $a_5$ and $a_6$.  We find that the either or both supernovae data
  sets have wide marginal distributions for the the parameters $a_1$ and
  $a_2$ and thus does not offer significant constraints.  Subsequently, we
  will not list them in the table below.  As well, we find that the R04 data
  does not constrain $a_3$.  The SNLS data constrains only two parameters $a_4$ and
  $a_5$.  Together, both data sets constrain $a_3$, $a_4$, $a_5$ and $a_6$. 
  Unconstrained parameters are denoted by dashes in the table.  All
  uncertainties are at 2$\sigma$.  Notice that the R06+SNLS minimum $\chi^2$
  is greater in the 6 point model than in the 3 point model.  This is due to
  parameter space grid is more dense than in the 3 point model.  Computational
  time was too prohibitive to fill in the grid to the same density as was done
  for the 3 point model.}
\begin{ruledtabular}
\begin{tabular}{cccccc}
Data Set      & $\chi^2_{{\rm min}}$& $a_3$ & $a_4$ & $a_5$ & $a_6$ \\
\hline
R04   & 174 & --  & $-0.3^{+0.3}_{-0.5}$  & $-0.6^{+0.4}_{-0.4}$ &
$-0.9^{+0.5}_{-0.5}$  \\ 
SNLS  & 61 & -- &  $-0.5^{+0.5}_{-0.6}$ & $-0.6^{+0.4}_{-0.4}$  & --\\
R04+SNLS  & 237  & $-0.3^{+0.3}_{-0.6}$  & $-0.4^{+0.4}_{-0.6}$  &
$-0.6^{+0.4}_{-0.4}$  & $-0.9^{+0.6}_{-0.6}$ \\
R06 & 200  & $-0.2^{+0.2}_{-0.6}$ & $-0.1^{+0.1}_{-0.4}$ & $-0.7 \pm 0.4$&
$-0.3^{+0.3}_{-0.4}$ \\ 
R06+SNLS & 264  & $-0.2^{+0.2}_{-0.4}$ & $-0.3^{+0.3}_{-0.6}$ & $-0.7 \pm 0.2$ &
$-0.3^{+0.3}_{-0.4}$ \\
\end{tabular}
\end{ruledtabular}
\end{table}
Comparing these two tables, we find that there is little statistical
difference between the significance levels of the marginal 3 and 6 point cubic
splines.  For the 3 point splines, all of the parameters are constrained.
With the 6 point splines, the R04 data constrains 3 parameters, $a_4$, $a_5$
and $a_6$.   The SNLS data constrains 2 parameters, $a_4$ and $a_5$.  For the
R04+SNLS data, we find that $a_3$, $a_4$, $a_5$ and $a_6$ are constrained.  
With the introduction of 21 new supernovae observations at redshifts $z>1$, we
find that the no new additional spline parameters are constrained.  However,
the uncertainties in the parameters appear to have been significantly
decreased.  
Multiply constrained spline parameters is in contrast to most of the models
considered in section \ref{sec:phenmodela} where only 1 parameter was
constrained for each model.
This is not unexpected for the spline models since multiple
spline points constrain $w_{tot}$ in the regions where the data is most
sensitive.    
The fits obtained here are marginally better than that which is obtained for
the previously discussed models.  

\begin{figure}

$\begin{array}{cc}
\includegraphics[width=3in]{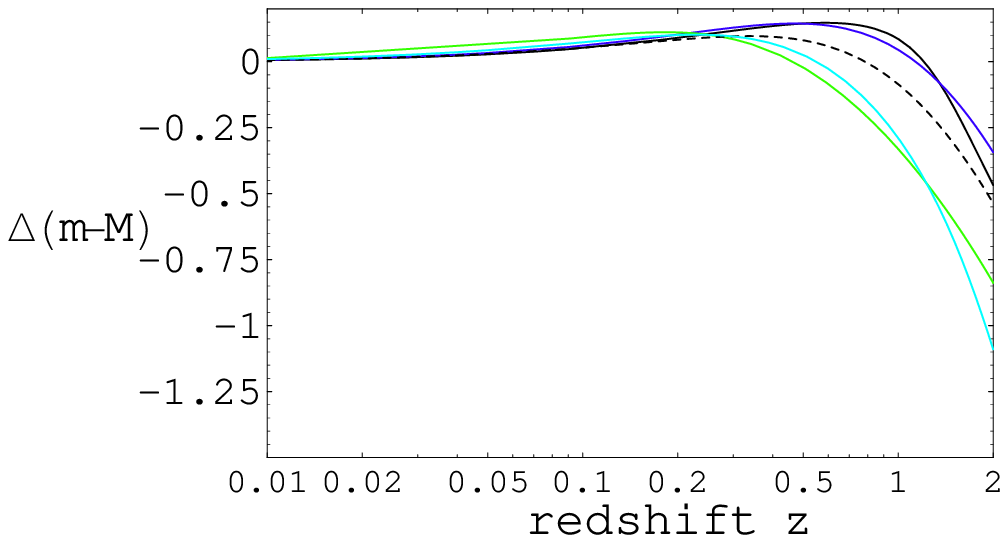} &
\includegraphics[width=3in]{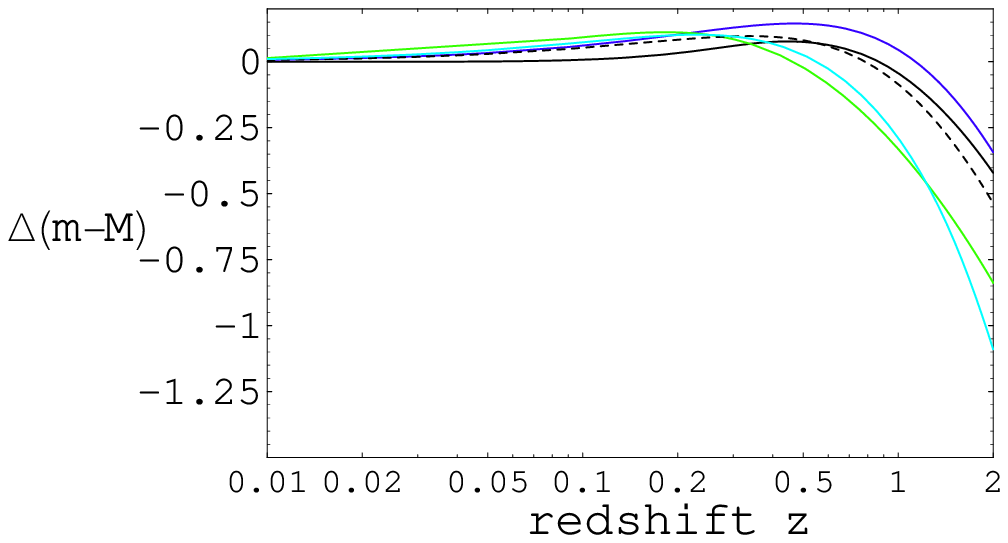} 
\end{array}$
\caption{(color online) A plot of the Distance Modulus $\Delta(m-M)$ versus
  redshift z for the R06+SNLS best fit $w_{tot}$ for the 3 and 6 point
  splines. The dashed line corresponds to the LCDM model with $\Om=0.27$.  The 
  green curve corresponds to the Linder 2.0 model with parameters $w_0=-2.25$
  and $w_1=0.3$.  The blue curve corresponds to the best fit VCDM model from
  \cite{Caldwell05} with $\Om=0.45$.  The 6 point spline model appears to be
  midway between the LCDM and VCDM models.  The 3 point spline most closely
  resembles the VCDM model.  From this result, it appears that the supernovae data
  favors models with a stronger transition to dark energy domination than that
  predicted with the LCDM and other models.  
\label{fig:spline2}}
\end{figure}

The supernovae data is given in terms of the difference between the apparent
magnitude ($m$) and the absolute magnitude ($M$).   A plot of this difference
modulo an open vacuum ($\Delta(m-M)$) versus redshift for the 3 and 6 point
marginal estimate obtained from the R06+SNLS data fit is given in figure
\ref{fig:spline2} along with the VCDM model, the LCDM model and the hyperbolic
model.  
Going from the 3 point marginal spline to the 6 point marginal spline in this
figure, we find that there is convergence to the marginal curve of the VCDM model.  
This implies that as the spline model favors a strong transition to dark
energy domination occurring at a redshift $z \approx 0.8$ which is more like
the transition coming from the VCDM model.  This is indicated
by the larger $\Delta(m-M)$ in figure \ref{fig:spline2} which corresponds to
greater observed dimming of the supernovae.  Weaker transitions (like
dark energy stemming from a non-zero cosmological constant) would generate
brighter supernovae and those correspond to a smaller $\Delta(m-M)$.  Thus,
the supernovae data appear to favor models which have greater transition than a 
cosmological constant.  

\begin{figure}
$ \begin{array}{ccc}
\includegraphics[width=2in]{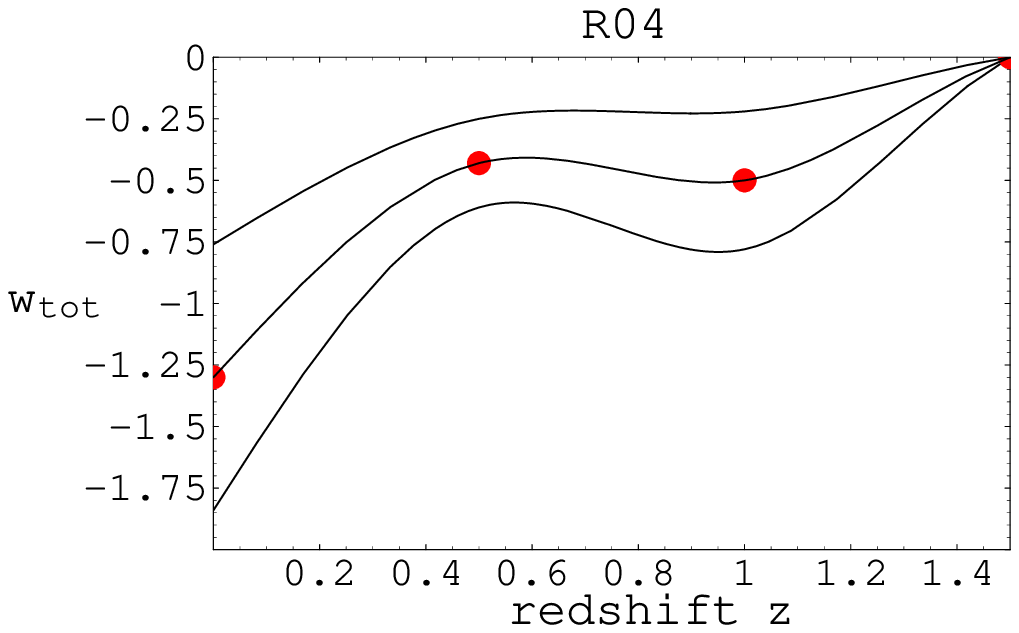} &
\includegraphics[width=2in]{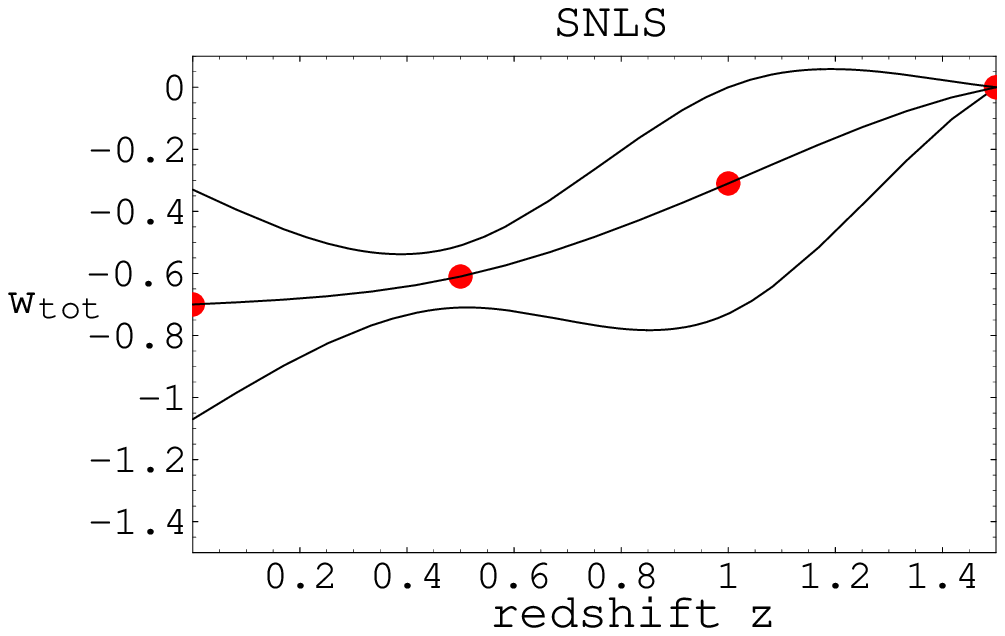} &
\includegraphics[width=2in]{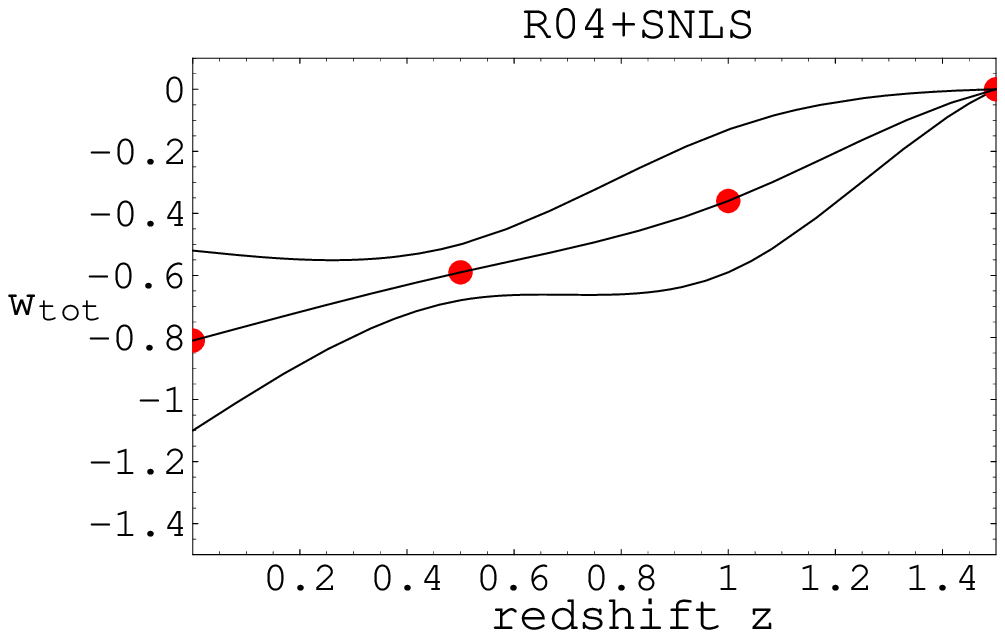} \\
\includegraphics[width=2in]{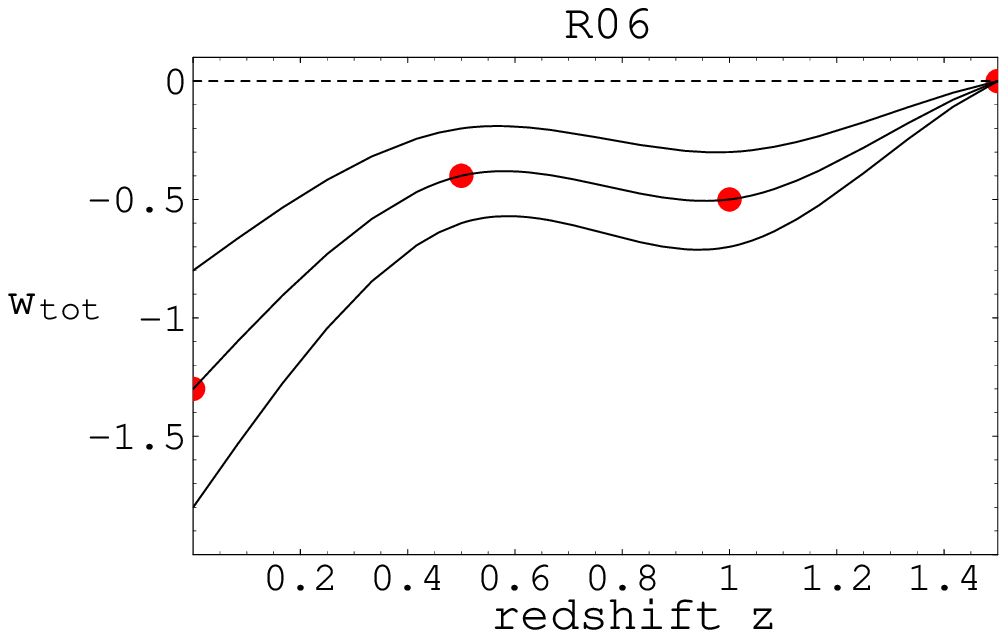} &
\includegraphics[width=2in]{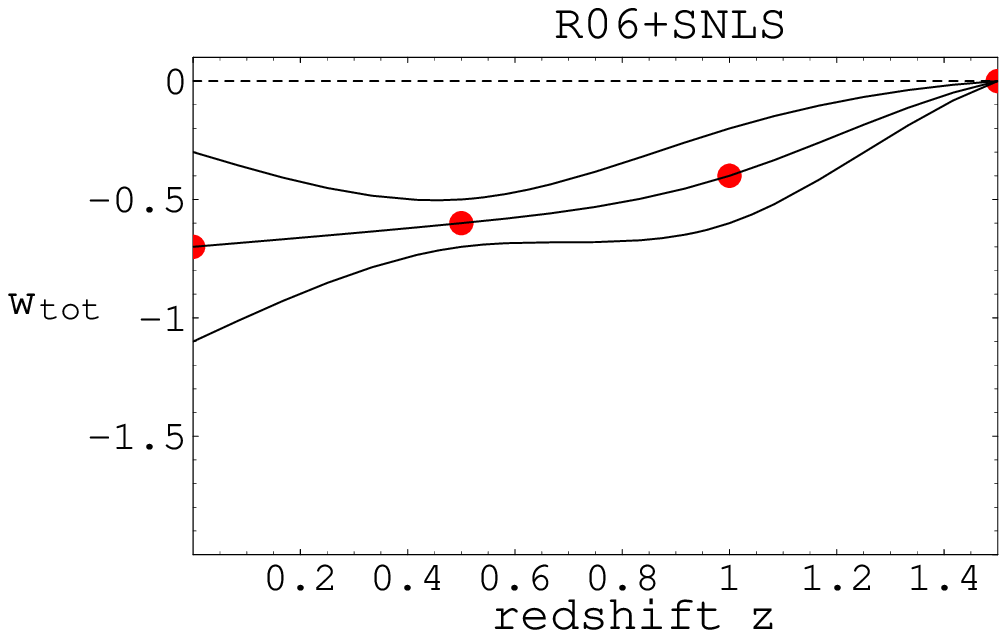} & 
\\
\includegraphics[width=2in]{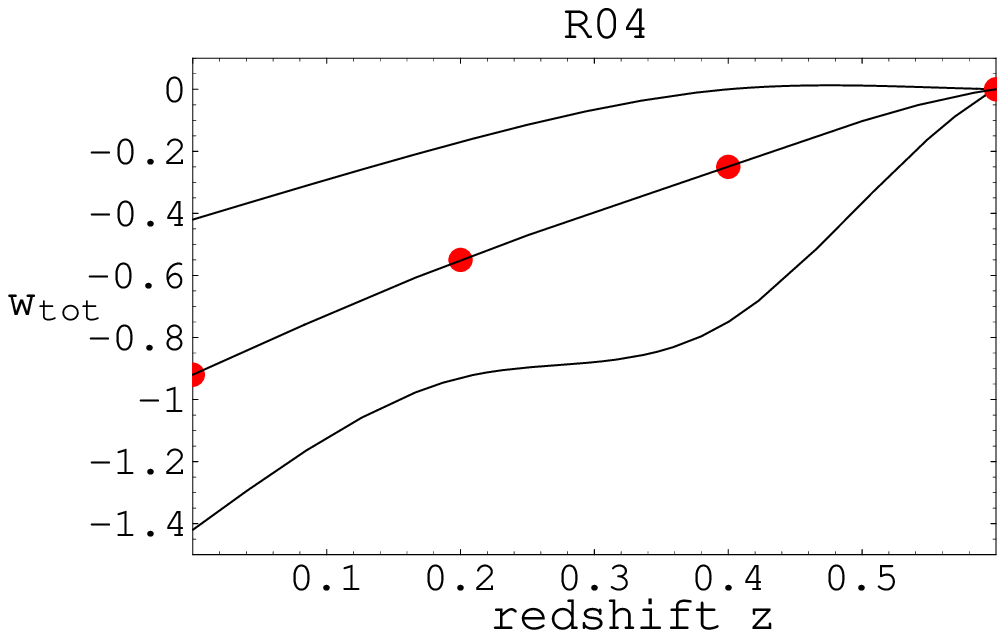} &
\includegraphics[width=2in]{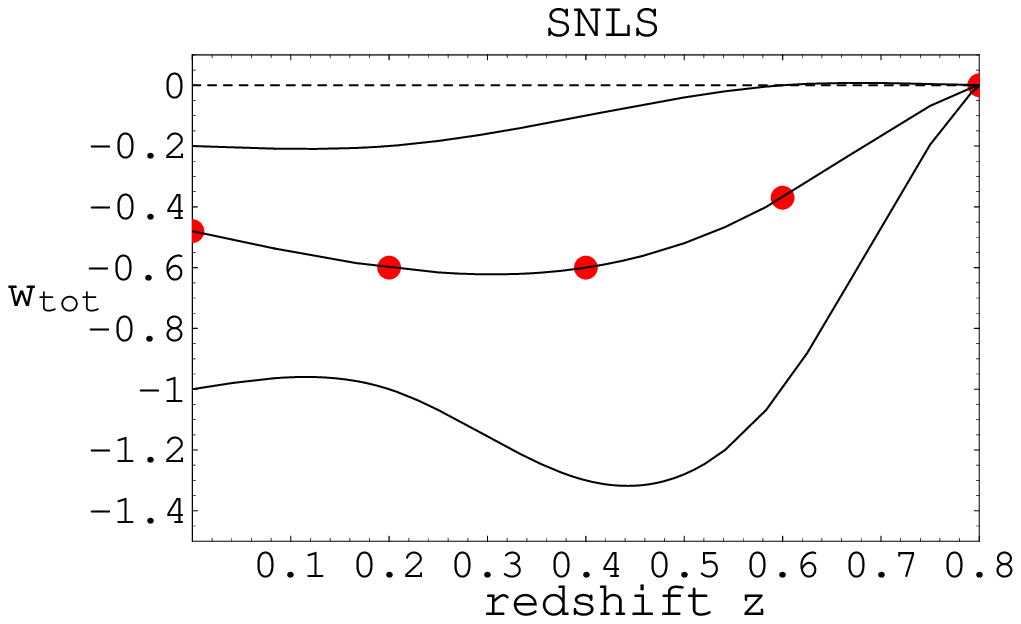} &
\includegraphics[width=2in]{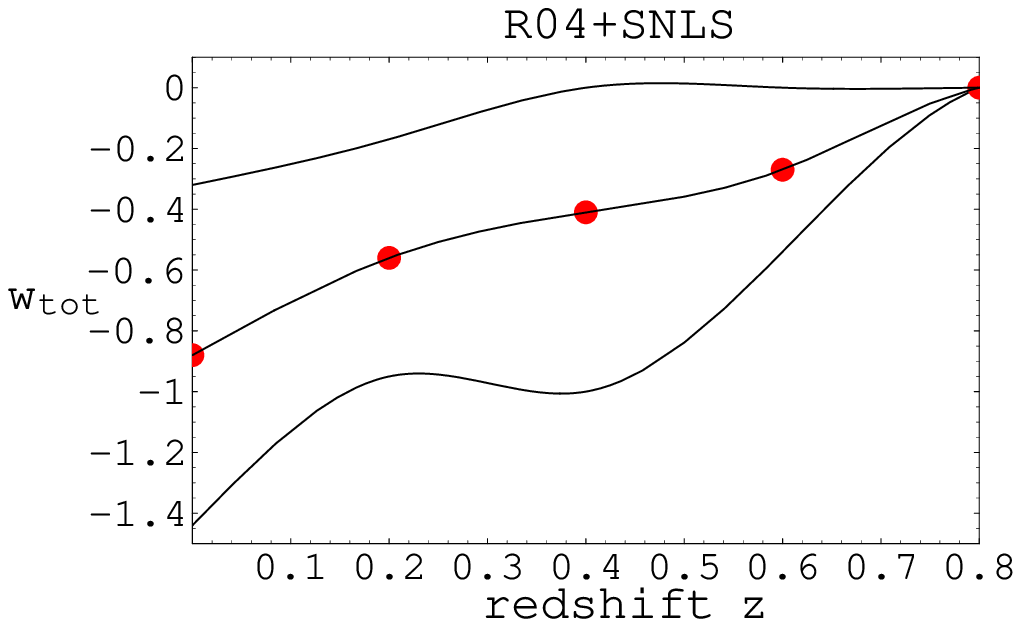} \\
\includegraphics[width=2in]{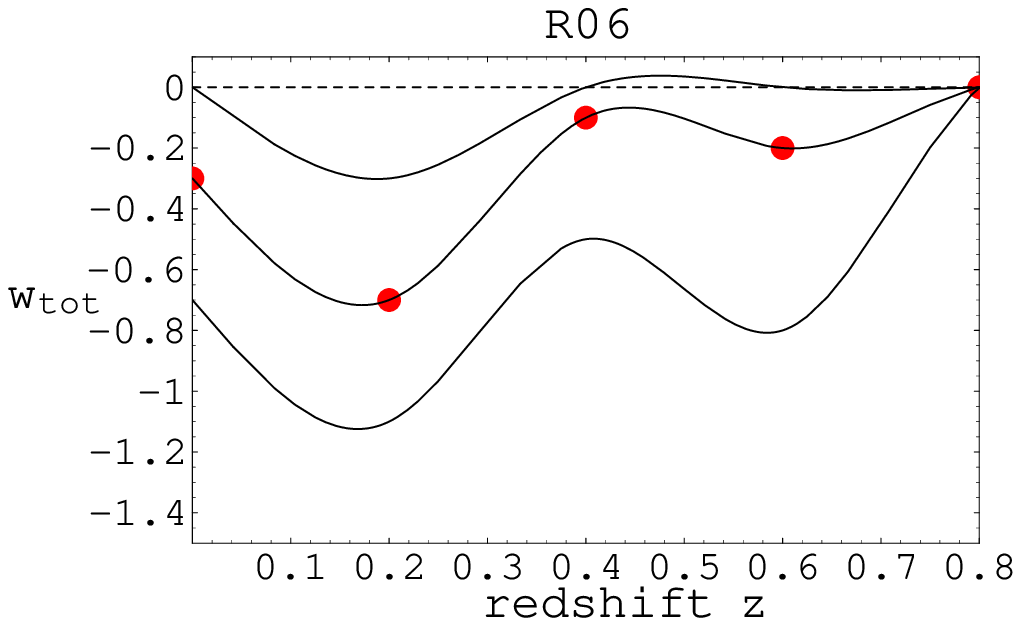} &
\includegraphics[width=2in]{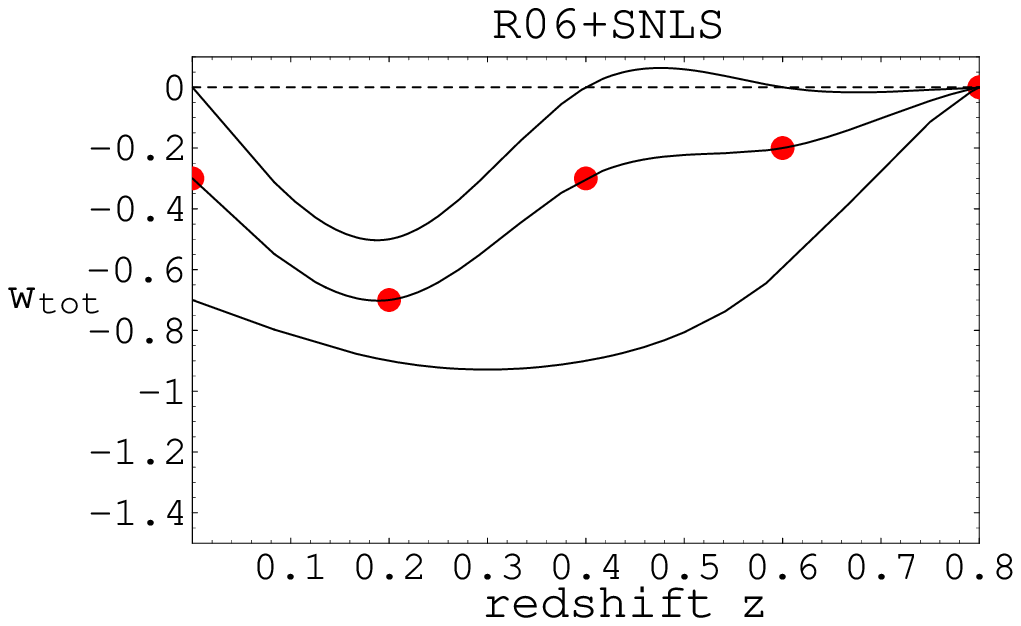}  & 
\end{array}$
\caption{(color online) A plot of the marginal estimates and 2$\sigma$
  confidence limits of $\wt$ for the 3 and 6 point
  spline models.  The results are for the R04, SNLS, R04+SNLS, R06 and
  R06+SNLS data sets.  The top two rows corresponds to the 3 point spline and the
  bottom two rows corresponds to the 6 point spline.  The splines show a different
  functional behavior of $\wt$ than that which was considered in the models
  from section \ref{sec:phenmodela}.  The functions here can have inflection,
  local maxima and minima.  From these plots, we find that the non-monotonic
  behavior in the dark energy occurs as high significance level.  Also, it is
  apparent that the most recent data release of R06 has increased this level.  
\label{fig:spline1}}
\end{figure}

A plot of the marginal
estimate of $w_{tot}$ and 2$\sigma$ confidence limits obtained by fitting to
the data is given in figure \ref{fig:spline1}.  
Also, this figure exhibits the
non-monotonic functional behavior of the splines.  Thus, these spline models
indicate that the data permits different functional behavior than that which
was considered in section \ref{sec:phenmodela}. 
Many of the spline models shown in figure \ref{fig:spline1} have
equations of state $\wt$ which experience non-trivial bouncing behavior 
in the dark energy dominated epoch at a redshift of $z \approx 0.5$.  
Accounting for parameter covariances, we find that such  bounces in $\wt$ are
within 1$\sigma$ confidence for each data set.  
For the LCDM \cite{Riess04,SNLS}, VCDM \cite{Caldwell05}
or the hyperbolic model.
we find that only one parameter is constrained (again assuming spatial
flatness) and each of these models have a monotonically decreasing equation of
state for the dark energy and consequently $\wt$.  Also, these models assume
that $\wt$ asymptotes to a constant value.
While the splines do have significant overlapping of confidence regions with
the other dark energy models from section \ref{sec:phenmodela}, the resulting
$\wt$ are not constrained to be monotonically decreasing.  

In \cite{Linder03}, model 2.0 has a monotonically decreasing equation of state
with two free parameters ($w_0$ and $w_a$) which are found to be
constrained.    
However, as mentioned previously this model has an asymptotic divergence in
its dark energy equation of state, i.e. it does not asymptote to a constant.  
Thus for any proposed dark energy model, the supernovae data can 
generically can constrain at most 4 degrees of freedom depending on the
assumptions of the underlying model.  
For the 6 point spline, constraints for
the R06+SNLS data given in table \ref{tab:6ptmarg} show 4 parameters being
constrained.  This represents an improvement over the R04 data sets.  It
is expected that these constraints will be reduced further with the upcoming
ESSENCE data \cite{essence}.  

\begin{figure}
$\begin{array}{cc}
\includegraphics[width=3in]{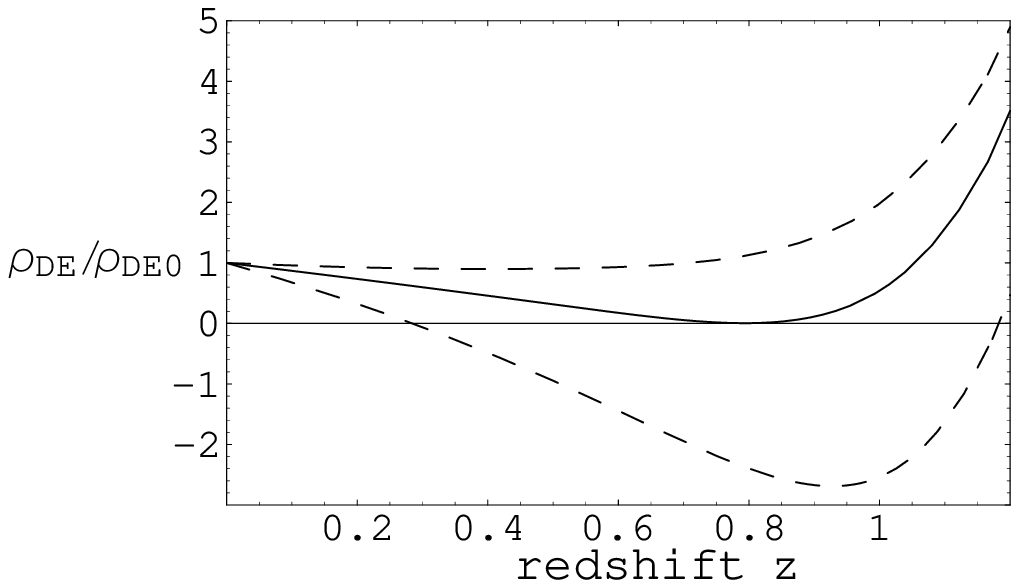} &
\includegraphics[width=3in]{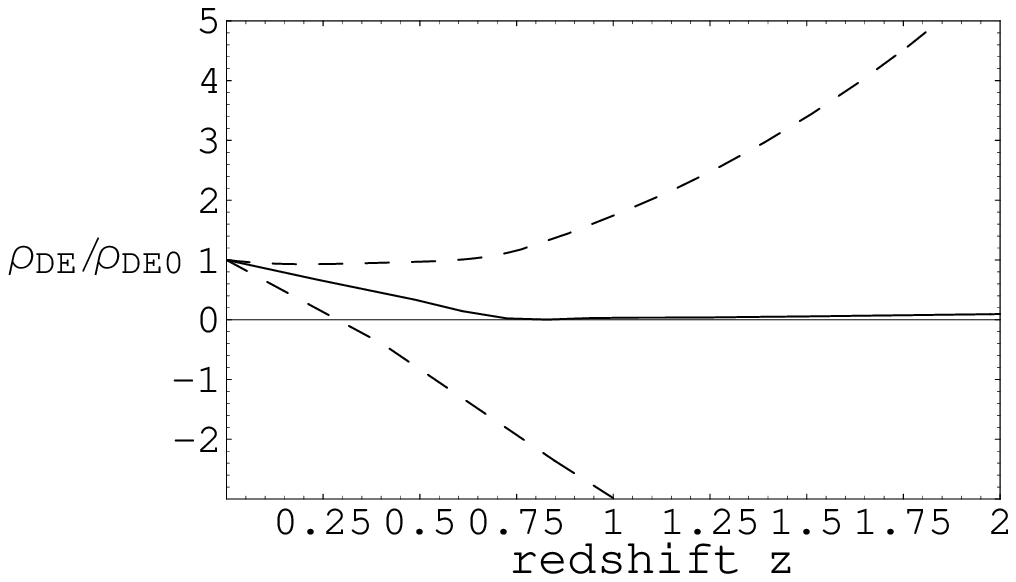} 
\end{array}$
\caption{Plots of the ratio of dark energy density
  ($\rho_{DE}$)to present day value ($\rho_{DE 0}$) for the 3 point spline
  (right) and 6 point spline (left) best fit models. For
  $z>1$, both spline models behaves as ordinary non-relativistic matter with
  $\wt \approx 0$.  At about $z \approx 1$, the dark energy under goes a
  transition which causes it to differ from ordinary non-relativistic matter.
  This transition is the source of the dark energy dominated epoch in both of
  the spline models.  The curves going from the top curve to the lowest curve
  correspond to $\Om=0.30,0.44,0.60$ respectively.  For $\Om>0.44$, the ratio
  goes negative.  This implies that the dark energy density can violate the
  Weak Energy Condition (WEC), but the total energy density never violates
  this condition.  Notice for the solid black curve ($\Om=0.44$) that the
  energy density vanishes for $z \approx 0.8$.  In the case of the 6 point 
  spline, we find that the dark energy is approximately 0 until about a
  redshift of 0.9 and rises significantly at late times.  
  This is dynamically very similar to the Sudden Gravitational Transition 
  models\cite{pkv2003,Caldwell05}.  
  \label{fig:spline3}}
\end{figure}

As with the hyperbolic model, consider $\rho_{tot}=\rho_{m}+\rho_{DE}$ and
determine the properties of the dark energy coming from the spline model that
are present in the observational data.  
Substituting this expression for $\rho_{tot}$ in \eqref{eq:rhotot} and solving
for $\rho_{DE}$ gives
\beq \label{eq:sprhode}
\frac{\rho_{DE}}{\rho_{DE0}}=\frac{1}{1-\Om} {\rm Exp} \left[ 3 \int_0^z \frac{dz'}{1+z'}
 (1+w_{tot}(z')) \right] -\frac{\Om}{1-\Om} (1+z)^3\;.
\eeq
A plot of this expression is shown in figure \ref{fig:spline3} for 
$\Om=0.30, 0.45$ and $0.60$ assuming the 3 and 6 point spline models.  In this
figure, the dark energy is behaving similarly to the dark energy from the
hyperbolic model.  Here, the dark energy at $z>1$ is behaving as matter until
reaching a minimum at $z \approx 0.8$.  From there, its evolution begins to
deviate from matter and this results in a decreasing $\wt$ and gives rise to
the effects that are attributed to dark energy.  This is not unlike what
occurs with previously 
considered models of dark energy.  From figure \ref{fig:spline3}, we see that
the decomposition of  $\rt$ implies a negative $\rho_{DE}$ for $\Om>0.45$.
This implies that the dark energy density could violate the Weak Energy
Condition (WEC), \cite{Visser,Wald,Hawking,Santos06}.  However, the total
energy density of the universe is always positive and thus WEC is not
violated.  This could lead to subtle effects on the vacuum solution of
Einstein Equations similar to those discussed in \cite{Olmo05} for $f(R)$
gravities and might be  worthy of some future study. It is worth pointing out
that the critical value $\Om=0.45$ corresponds to the marginal estimate of the
VCDM model.   

All of the models discussed in
section \ref{sec:phenmodela} had monotonically decreasing $\wt$'s.  
There are as of yet no models which
possess this bouncing characteristic on scales of order unity in
redshift.  In \cite{Parker04}, they show that the VCDM model possesses a
bouncing characteristic with respect to the order parameter $R^2$ around the
time of transition to dark energy domination.  However, they found that this
behavior experiences a rather rapid exponential decay in the dark energy epoch
and asymptotes rather rapidly to a constant value on a time scale much smaller
than the hubble time.  
  
\begin{figure}
$\begin{array}{cc}
\includegraphics[width=2in]{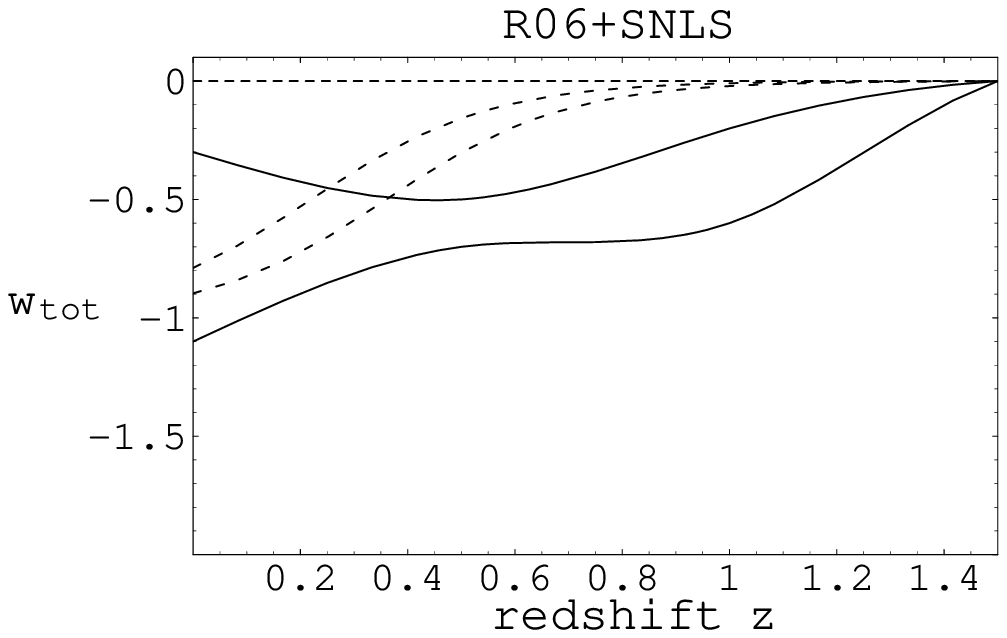} &
\includegraphics[width=2in]{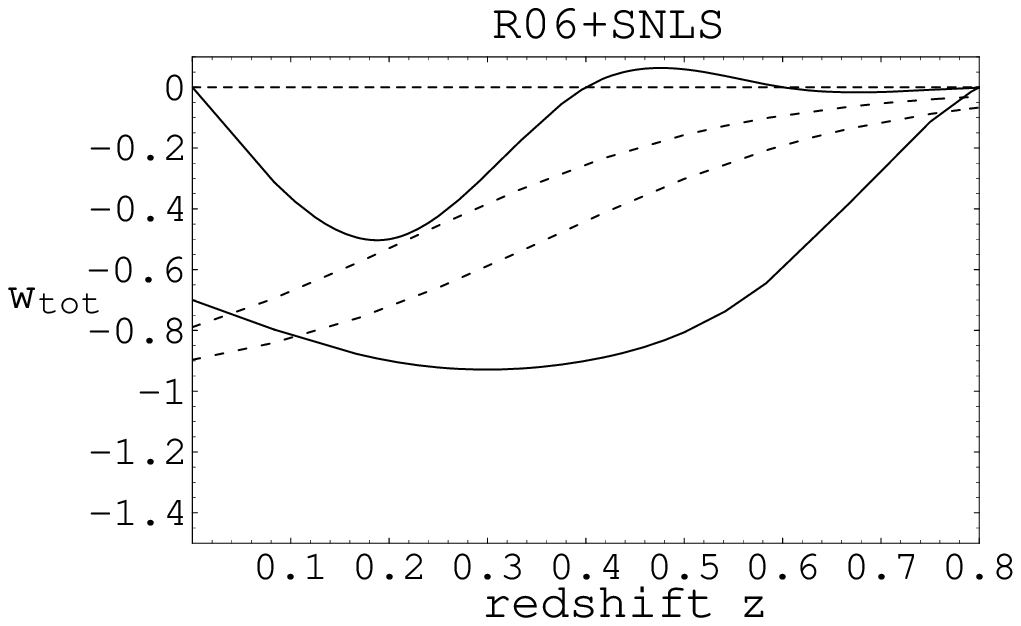} 
\end{array}$
\caption{(color online) A plot of the 2$\sigma$ confidence limits for
  $\wt$ obtained from the 3 and  6 point cubic spline models (top and lower
  plots respectively).  
  The solid and dashed black curves are the best fit
  and 2$\sigma$ confidence limits for the hyperbolic model. 
  The blue curves correspond to the VCDM model best fit
  (solid curve) and the 2$\sigma$ confidence limits (dashed curves).  For this
  plot, if no constraint was obtained from the data fit then all values
  between 0 and -1.4 are possible.  For sake of easy of plotting,
  unconstrained parameters were assigned the value of 0.  Notice,
  that the regions of greatest intersection of the models correspond to
  precisely the region shown in figure \ref{fig:compplot} around redshifts of
  $z \approx 0.3$ for the 3 point splines and $z \approx 0.8$ for the 6 point
  splines.   
\label{fig:spline4}}
\end{figure}

Despite the dynamical differences in $\wt$ between the dark energy models
consider in this and the previous sections, we
show in figure \ref{fig:spline4} that the various cosmological models have
significant overlap in or are very close to the 2$\sigma$ limits obtained
from the spline analysis for both 3 and 6 point splines.  
For the SNLS 3 point spline fit, the very tight
constraints on $a_2$ result in the greatest difference of the regions.
However, the 6 point spline more than encompasses all of the confidence
regions of the other models.  This is more than indicative of the potential
constraints on cosmological models.  As we know from the $\chi^2_{min}$'s from
tables \ref{tab:3ptmarg} and \ref{tab:6ptmarg} that there is little
significance difference between the 3 and 6 point spline models.  Thus, if one
is interested in model independent constraints for dark energy arising from
these two spline models, then the constraints derived from the 6 point spline
are a more fair estimate.   

\section{Discussion} \label{sec:disc}
Now lets consider the implications of the two models considered above and
compare them with the recent supernovae results of R06.  There they
published 21 new supernovae at high redshifts $z>1.0$.  The fitting to the
data utilizes the Monte Carlo Markov Chain algorithm from Lewis and
Briddle\cite{lewis02}.  They are able to draw significant constraints on the
values of the present day dark energy density assuming it arises from  a
non-zero cosmological constant.  Included in their analysis are potential
non-monotonically decreasing dark energy equations of state.  They say that
these models are 
ruled out due to Occam's Razor stemming from the greater number of model
parameters.  

We aim to show that the implications of this analysis are not as
clear as they may first appear.  There are certain assumptions on the total equation
of state of the cosmological fluid which one must assume in order to have a
LCDM model in the context of $\wt$.  The resulting implications for a relative
model comparison make direct comparisons quite difficult to analyze
quantitatively.  However with a few simplifying assumptions, a model
comparison that includes all of the models priors can be made.
In the
following analysis, we want to calculate the relative probability of the LCDM
versus a 3 point spline model for the R06 data.  We will compute the Occam
Factors and relative likelihoods associated with each model given the recent
supernovae data, see \cite{Sivia} for a good example.  It will be shown that
the supernovae data permits us to constrain at most 3 or at most degrees of
freedom of any perspective cosmological model.  In the context of this paper,
we will take the perspective of the total equation of state.  This is
important since we are trying to constrain the potential behavior of dark
energy and determine the degrees of freedom associated with dark energy.

Understanding how to compare relative probabilities of the two models begins
with Bayes' Theorem and multiplication property of probabilities,
i.e. respectively 

\beq
P(X|Y,I)=\frac{P(X|I)}{P(Y|I)} P(Y|X,I) \label{eq:bayes} 
\eeq

\beq
P(X,Y|I)=P(X|Y,I) P(Y|I) \label{eq:product} \;,
\eeq
where P stands for the probability, Y and X are two propositions, I is a
universal set.  For an arbitrary cosmological model, X stands for the model
and Y represents the superovae data (SNLS and/or R06).

For the LCDM model, we want to compute $P(LCDM|SN,I)$.  This model has certain
assumptions or propositions that are built in which distinguish it from the Spline models.
We will analyzed these in terms of $\wt$.  The first proposition is
monotonicity of $\wt$, i.e. $\wt(z_1)<\wt(z_2)$ for $z_1<z_2$.  This
proposition will be labeled hence forth as A and the apriori probability function is
$P(A|I)$.  Secondly, $\wt$ asymptotes to a constant ($-1$ for the LCDM model).
This proposition will be labeled B and its prior probability function is given
by $P(B|I)$.  
Any calculations of relative model probabilities must factor in the apriori
probability of these propositions when one compares the LCDM model to a model
without these assumptions.  We will find that the associated prior probability
functions associated with these propositions will be included in the Occam
Factor.   

From \eqref{eq:bayes}, the function $P(LCDM|SN,I)$ can be written
\beq \label{eq:lcdm1}
P(LCDM|SN,I)=\frac{P(LCDM|I)}{P(SN|I)} P(SN|LCDM,I) .
\eeq
Now, we can rewrite the last term on the right hand of this equation in terms
of propositions A and B as follows
\beq \label{eq:lcdm2}
P(SN|LCDM,I) \equiv P(SN|A,B,I) .
\eeq
Rewriting $P(LCDM|I)$ using \eqref{eq:product} and substituting this and
\eqref{eq:lcdm2} into \eqref{eq:lcdm1} gives
\beq \label{eq:lcdm3}
P(LCDM|SN,I)=\frac{P(A|I) P(B|I)}{P(SN|I)} P(SN|A,B,I)\;,
\eeq
where the last term on the right hand side is proportional to the likelihood
function that is used to compare predictions of the LCDM model to the
supernovae data.  Also, we have assumed that $P(A|I)$ and $P(B|I)$ are not
apriori correlated in anyway.

We are not quite at the LCDM model.  Refinement of the last term on the right of
\eqref{eq:lcdm3} is required.  Proposition B states that the model being
considered asymptotes to a constant.  For the LCDM model, this is assumed to
be $-1$.  In \cite{Caldwell02}, they considers deviations from this value and
even makes it a free model parameter. The only difference between these two
asymptotic models are the prior distributions functions for the constant,
labeled c hereafter.  Computing the marginal likelihood over all values of c,
\eqref{eq:lcdm3} becomes 
\beq \label{eq:lcdm4}
P(SN|B,A,I)=\int_{{\rm D}} P(SN|c,B,A,I) P(c|B,I) dc\;,
\eeq
where {\rm D} is the domain of c and $P(c|B,I)$ is the prior probability
distribution function for c.  The LCDM model assumes that this function is
given by $P(c|B,I) \propto \delta(c+1)$.  In \cite{Caldwell02}, this
assumption is removed.  

Here, we are concerned solely with the LCDM model and its marginal probability
given the supernovae data.  With the prior on c, this leaves the LCDM model
with one free parameter, $\Lambda$. So, \eqref{eq:lcdm4} can be written in
terms of this parameter and gives 
\beq \label{eq:lcdm5}
P(SN|B,A,I)=\int_D P(SN|\Lambda,c=-1,B,A,I) P(\Lambda|c=-1,A,B,SN,I) d \Lambda \;,
\eeq
where $P(\Lambda|c=-1,A,B,SN,I)$ is the prior probability distribution
function for the free model parameter $\Lambda$.  So, combining
\eqref{eq:lcdm5} and \eqref{eq:lcdm3} gives
\beq \label{eq:lcdm6}
\begin{array}{l}
P(LCDM|SN,I)=\frac{P(A|I) P(B|I)}{P(SN|I)} \\
 \\ 
\int_{D} P(SN|\Lambda,c=-1,B,A,I) P(\Lambda|c=-1,A,B,SN,I) d \Lambda\;.
\end{array}
\eeq
This expression is the probability of the LCDM model given the supernovae data
alone.  If we want to compare the LCDM model to any other model.  This is the
term that we need to use.  Now, lets do the analogous calculation for the
spline model.  

Lets assume for the sake of simplicity the 3 point spline.  We want to compute
the probability of the model given the supernovae data ($P(Sp|SN,I)$) as we
did above for the LCDM model. To do this, we follow a similar argument as
before.  Using \eqref{eq:bayes}, this probability can be written 
\beq \label{eq:spline1}
P(Sp|SN,I)=\frac{P(Sp|I)}{P(SN|I)} P(SN|Sp,I)\;.
\eeq

We can express $P(SN|Sp,I)$ as an integral over the parameter space
\beq \label{eq:spline2}
P(SN|Sp,I)=\int_{V}  P(SN,a_1,a_2,a_3|Sp,I) da_1 da_2 da_3 \;,
\eeq
where $a_1$, $a_2$ and $a_3$ are the model parameters and V is the volume of
the 3-d parameter space.  Using \eqref{eq:product}, this can be written as
\beq \label{eq:spline3}
\begin{array}{l}
P(SN|Sp,I)= \\ \\
\int_{V} P(SN|a_1,a_2,a_3,Sp,I) P(a_1|Sp,I) P(a_2|a_1,Sp,I) 
P(a_3|a_1,a_2,Sp,I) da_1 da_2 da_3
\end{array} \;,
\eeq
where $P(SN|a_1,a_2,a_3,Sp,I)$ is the likelihood function; $P(a_1|Sp,I)$,
$P(a_2|a_1,Sp,I)$, $P(a_3|a_1,a_2,Sp,I)$ are the parameter prior probability
functions.  There is no reason prior to fitting to the supernovae data to
assume that the parameters are correlated.  So, assuming apriori that the
parameters are uncorrelated gives
\beq \label{eq:spline4}
P(SN|Sp,I)=\int_{V} P(SN|a_1,a_2,a_3,Sp,I) P(a_1|Sp,I) P(a_2|Sp,I)
P(a_3|Sp,I) da_1 da_2 da_3\;.
\eeq
For sake of simplicity, lets assume uniform priors for the prior
probability functions in \eqref{eq:spline4}.  This equation can be written
\beq \label{eq:spline5}
P(SN|Sp,I)= P(a_1|Sp,I) P(a_2|Sp,I)
P(a_3|Sp,I) \int_{V} P(SN|a_1,a_2,a_3,Sp,I) da_1 da_2 da_3\;.
\eeq  

If we assume that the prior probability distribution function
$P(\Lambda|c=-1,A,B,SN,I)$ in \eqref{eq:lcdm6} is a uniform distribution as
well, then we can pull it out front of the integral.  
From \eqref{eq:lcdm6} and \eqref{eq:spline5}, the relative probability of the
LCDM model to the Spline model given the supernovae data is
\beq \label{eq:relprob}
\frac{P(LCDM|SN,I)}{P(SN|Sp,I)}= OF 
\frac{\int_{{\rm D}} P(SN|\Lambda,c=-1,B,A,I)
 d \Lambda} {\int_{V} P(SN|a_1,a_2,a_3,Sp,I) da_1 da_2
da_3}\;,
\eeq
where $OF$ stands for the Occam Factor\cite{Sivia} given by the ratio of the priors out
front on the integrals and has the form 
\beq \label{eq:of} 
OF=\frac{P(A|I) P(B|I) P(\Lambda|c=-1,A,B,SN,I) }{P(a_1|Sp,I) P(a_2|Sp,I)
P(a_3|Sp,I)}\;.
\eeq
When fitting to the data, we approximate the posterior distribution
function for a general cosmological model given the supernovae data as 
\beq
P(SN|\vecal,{\rm Model},I)d \vecal={\rm Exp}\left[-\chi^2(\vecal)/2 \right]\;,
\eeq
where $\vecal$ represents the model parameters and $\chi^2$ defined in
\cite{R06}.  

The ratio integrals in \eqref{eq:relprob} are the relative marginal likelihood
of the LCDM and Spline models that is determined by fitting to the data.  
If the relative probability only involved the ratio of the integral terms,
then obviously models with more parameters would be favored over simpler
models.  The point of the Occam Factor is to counteract this ratio and keep
over complex models from being favored simply due to their greater number of
degrees of freedom.  This is the idea considered in
\cite{R06} when comparing more complex spline models (with 3 or more
parameters) relative to the LCDM model.  However, this is not the only
consideration when determining the relative likelihood.  The LCDM model has
prior probabilities in the Occam Factor that involve the assumptions A and B
above.  It is difficult for us to quantitatively determine precisely what
these probabilities are.  

As a simple model, consider that these priors are unity and assume
that both models will initially begin as Standard Cold Dark Matter models
(i.e. no dark energy).  We expect from the WMAP and other cosmological data
that $\Omega_{\Lambda} \approx 3/4$.  Since, we are assuming uniform priors for
the parameters of both models then it reasonable to expect from the tight WMAP
that the limits of$\Omega_{\Lambda}$ is bounded by the interval [0,0.9].  For
the spline models, we assumed uniform priors $a_1$, $a_2$ and $a_3$ from the
interval [0,-1.4].  Plugging this into the 
\eqref{eq:of} gives $OF \approx 3$.  The majority of the probability
associated with each marginal distribution function is located near the
minimum in $\chi^2$.  So, we can approximate the ratio of the
marginal likelihood distribution functions as 
\beq \label{eq:margratio}
\frac{\int_{D} {\rm Exp}[-\chi^2(\Lambda)/2] d\Lambda}{\int_{V} {\rm
    Exp}[-\chi^2(a_1,a_2,a_3)/2] da_1 da_2 da_3 } \approx \frac{{\rm
    Exp}[-\chi^2(\Lambda_{0})/2]}{{\rm
    Exp}[-\chi^2(a_{10},a_{20},a_{30})/2]}\;,
\eeq
where the 0 subscript denotes best fit value, and we have set the ratio of the
volumes of parameter spaces to 1.  We found for the R06 data that the spline
model has a $\chi^2$ which is about 4 lower than the LCDM model.  
Thus, \eqref{eq:margratio} is approximately 0.2.  So despite the increase
number of model parameters associated with the 3 point spline models, the data
still appears to favor dark energy associated with splines over the LCDM
model. Furthermore, we can interpret the implications of the resulting fit and  
conclude that there is some significance to the idea that assuming 
propositions A and/or B is too restrictive.
In this simple model, we have $OF \approx 3$ and the ratio of the marginal
likelihoods is given by $e^{-2.0}$.  Now lets turn the argument above around
and compute the probability of the A and B being true given the supernovae
data.  So, solving \eqref{eq:relprob} and \eqref{eq:of} for $P(A|I) P(B|I)$
and substituting in the values from the previous paragraph gives 
\beq
P(A|I) P(B|I) \approx \frac{e^{-2.0}}{3} \left( \frac{1.4^3}{0.9}
\right) \approx 0.37\;.
\eeq
Thus in the approximation of uniform priors for the LCDM model, we
find that there is significance to the idea that the physics is more dynamical
than what occurs in the LCDM model.  This does not say that the LCDM 
model is to disregarded, but it does say that at present the SN data can not
distinguish between the spline models (multiple parameters and
non-monotonic behavior) and the LCDM model (one parameter and monotonicity).
In fact, the spline models can serve as a test of some of the underlying
assumptions of the LCDM model.  This leaves open the door to some of the more
exotic models of dark energy discussed in section \ref{sec:intro} which are
dynamically very different than the LCDM model.  One could propose to increase
the number of parameters, e.g. introducing spatial curvature to improve the
fit.  In \cite{komp04,Caldwell05}, non-flat models are shown to improve the
fit but only marginally.  For the LCDM model, this does lead to a significant
expansion of the 
likelihood contours due to the high correlation between $\Lambda$ and the
curvature parameter, $\Omega_{k0}$. However, this does not really affect the results
discussed above and does not account for the WMAP data which places very
tight constraints on curvature\cite{Bennett06} (similar correlations were
found for the curved VCDM models in \cite{Caldwell05,komp04}).  

Lets now compare the predictions of the $\wt$ models with the predictions of
WMAP\cite{Bennett06}.  To do this, we will use the results of Wang and
Mukherjee\cite{Wang06}.  They derive model independent constraints on the
shift parameter $R=\sqrt{\Om}\int_{0}^{z_{rec}} dz/E(z)$ where
$E(z)=H(z)/H_0$, $\Om$ is the present day matter density and $z_{rec}$ is the
redshift of recombination ($\approx 1100$).  The models that we
are considering in this paper do not have a clearly defined value of $\Om$.
However, one point that both models have in common and would be
indistinguishable would be for $z \gg z_j$ where $z_j$ is the redshift of
transition to dark energy domination.  In this limit, both models are
dominated by non-relativistic matter.  For the LCDM model, this means that
$H^2(z)/H_0^2 \approx \Om (1+z)^3$.  

For the spline models, $z \gg z_j$ means that $\wt \approx 0$.  Thus,
\eqref{eq:phen00ee} and \eqref{eq:rhotot} can be written  
\beq
E^2(z)= {\rm Exp} \left( 3 \int_0^{z_j} \frac{dz'}{1+z'} (1+\wt (z) )\right)
\frac{(1+z)^3}{(1+z_j)^3}\;.
\eeq
Comparing this result to the corresponding result for the LCDM model, we can
define a dimensionless matter density which evolves in a similar fashion
in the spline model as follows 
\beq \label{eq:splom}
\Om = (1+z_j)^{-3}  {\rm Exp} \left( 3 \int_0^{z_j} \frac{dz'}{1+z'} (1+\wt
  (z) )\right)\;.
\eeq
With this definition, \eqref{eq:phen00ee} can be written as 
\beq
E(z)^2=\left\{
\begin{array}{ll}
\Om (1+z)^3 & {\rm for} \ z>z_j \\
\Om (1+z_j)^3 {\rm Exp} \left[- 3 \int_z^{z_j} \frac{dz'}{1+z'} (1+\wt (z)
  )\right]  & {\rm for} \  z<z_j
\end{array}\right\} \;. 
\eeq

Using the results from fitting to the supernovae,we find that $\Om=0.3 \pm
0.05$ at 2$\sigma$.  This range has significant overlap with the
LCDM estimate of $\Om=0.27 \pm 0.02$\cite{Bennett06}.  Using this results and
the above expression for the shift parameter, we get $R=1.8 \pm 0.1$ at
2$\sigma$.  In \cite{Wang06}, they find that $R=1.70 \pm 0.03$ at 1$\sigma$.   

In this section, we have compared model implications coming from the new
supernovae data to that of the 3 point spline model considered in this paper.
We have shown that one must be careful in the assumptions of any
parameterization of dark energy.  Also, we have shown that the most recent
WMAP CMB data can be consistent with models that have non-monotonically
decreasing total equations of state.  

\section{Conclusion} \label{sec:con}

We have shown in this paper, that the supernovae data offers very
tight constraints on $w_{tot}$.  This analysis assumed two very different 
cosmological models.  The first is a 2 parameter phenomenological model based
on the hyperbolic tangent.  The supernovae data gives parameters estimates of
$\alpha=0.29 \pm 0.08$ and $\beta=3.2 \pm 2.8$ at 2$\sigma$ for the R06+SNLS
data.  As with many other dark 
energy models, we find that only the parameter $\alpha$ (corresponding to the
time of transition) is significantly constrained leaving $\beta$ relatively
unconstrained.  By varying these parameters, we find that this model is able
to reproduce similar transitions to dark energy domination as several 
previously considered dark energy models.  Of these models, the
hyperbolic model best reproduces transitions like those coming from the
Sudden Gravitational Transition Models\cite{Caldwell05} and is more physically
attuned to the dark fluid model\cite{Arbey05}.   

The other model of dark energy that we considered was a cubic spline model of $\wt$.
This model supposes a prior matter dominated epoch for $z \gg 1$ and has the
potential of some non-trivial dark energy behavior for $z<1$.  
Two different uniform density of spline points were considered one with 3
spline points and 6 points from the interval $z \in [0,1]$.  
These models are similar to those considered in \cite{Wang04} but there they
were used to constrain the dark energy density.  
In this paper, we model the total equation of state $\wt$ and find that the
supernovae data has a very wide flexibility in the total equation of state.
As many as 4 spline points can be constrained by the supernovae data with a
minimum number of assumptions.  There is significant overlap of the 2$\sigma$
confidence regions between these models and the monotonically decreasing models
considered in section \ref{sec:phenmodela}.  
This is consistent with results presented in \cite{Wang04}.  
Also, the models permit the possibility that the
dark energy by itself can violate the Weak Energy Condition (WEC), but
together with matter there is no violation of this condition.  

We find that there is general agreement between the cosmological models
considered in this paper and the cosmological data coming from supernovae and
WMAP.  At present, there is not enough information to determine the precise
physical behavior of dark energy and thus we have little clue to its origins
and/or underlying physics.  This work suggests that our scope of investigation into dark
energy should not be limited to the LCDM model or models which are solely monotonically
decreasing.  The present data permits and even marginally favors models with
non-trivial dynamics. Perhaps, future observations with more accurate
observations would give increased knowledge about the dynamical behavior of
$\wt$.  We have shown that the update of R04 with the R06 data has reduced the
estimated parameters uncerainties by about 10\%.  The upcoming ESSENCE
project\cite{essence} results may provide some additional upper limits on this
type of behavior.  The proposed Supernova Acceleration Probe
(SNAP)\cite{SNAP} at roughly 2000 observed supernovae per year should be able
to distinguish between monotonic and non-monotonic $\wt$ through its
observation of the time evolution of the equation of state.   

Future work could include determining the precise solutions of the
perturbation equations which could potential constrain $\wt$.  This would
allow comparison with the growth of Large Scale Structure.  Examining the
subtle effects of these equations of state to post-Newtonian approximation may
lead to some constraints as well.

\acknowledgements
W. K. would like to thank Leonard Parker for useful comments and
discussions.  I would also like to thank the Physics Department at
the University of Louisville for their hospitality during the duration
of this project.  Also, a great thanks is owed to James T. Lauroesch for very
useful discussions and editing comments.  This work was partially supported by a
Research Initiation Grant from the University of Louisville.  This work is
dedicated to my late father, Joel T. Komp.  RIP dad.

\end{document}